\documentstyle[psfig]{mn2e}
\input epsf

\def\deg{\hbox{$^\circ$}}

\def\lesssim{\mathrel{\hbox{\rlap{\hbox{\lower4pt\hbox{$\sim$}}}\hbox{$<$}}}}
\def\gtrsim{\mathrel{\hbox{\rlap{\hbox{\lower4pt\hbox{$\sim$}}}\hbox{$>$}}}}

\begin{document}

\title[H$\alpha$ survey of the northern Galactic Plane (IPHAS)]
{The INT Photometric H$\alpha$ Survey of the Northern Galactic Plane (IPHAS)}
\author[Janet E. Drew et al. ]
{Janet E. Drew$^1$, R. Greimel$^2$, M. J. Irwin$^3$, A. Aungwerojwit$^4$, 
M. J. Barlow$^5$, 
\newauthor R. L. M. Corradi$^2$, J. J. Drake$^6$, B. T. G\"ansicke$^4$, 
P. Groot$^7$, A. Hales$^5$, 
\newauthor E. C. Hopewell$^1$, J. Irwin$^3$, C. Knigge$^8$, P. Leisy$^{9,2}$, 
D. J. Lennon$^2$, A. Mampaso$^9$, 
\newauthor M. R. W. Masheder$^{10}$, M. Matsuura$^{11}$, L. Morales-Rueda$^7$, 
R. A. H. Morris$^{10}$, 
\newauthor Q. A. Parker$^{12,13}$, S. Phillipps$^{10}$, 
P. Rodriguez-Gil$^{4,9}$, G. Roelofs$^7$, I. Skillen$^2$, 
\newauthor J. L. Sokoloski$^{6}$, D. Steeghs$^{6}$, Y.C. Unruh$^1$, 
K. Viironen$^9$, J. S. Vink$^1$, 
\newauthor N. A. Walton$^3$, A. Witham$^8$, N. Wright$^5$, 
A. A. Zijlstra$^{11}$, A. Zurita$^{14}$\\ 
$^1$Imperial College of Science, Technology and Medicine,
Blackett Laboratory, Exhibition Road, London,  SW7 2AZ, U.K.\\
$^2$Isaac Newton Group of Telescopes, Apartado de correos 321,
E-38700 Santa Cruz de la Palma, Tenerife, Spain\\
$^3$Institute of Astronomy, Cambridge University, Madingley Road, Cambridge,
CB3 OHA, U.K.\\
$^4$Department of Physics, University of Warwick, Coventry, CV4 7AL, U.K.\\ 
$^5$University College London, Department of Physics \& Astronomy, 
Gower Street, London WC1E 6BT, U.K.\\
$^6$Harvard-Smithsonian Center for Astrophysics, 60 Garden Street, 
Cambridge, MA 02138, U.S.A. \\
$^7$Afdeling Sterrenkunde, Radboud Universiteit Nijmegen, Faculteit NWI,
Postbus 9010, 6500 GL Nijmegen, The Netherlands\\
$^8$ School of Physics \& Astronomy, University of Southampton, Southampton,
SO17 1BJ, U.K.\\ 
$^9$ Instituto de Astrofisica de Canarias, 38200 La Laguna, Tenerife, Spain\\
$^{10}$Astrophysics Group, Department of Physics, Bristol University, 
Tyndall Avenue, Bristol, BS8 1TL, U.K.\\
$^{11}$ School of Physics and Astronomy, University of Manchester,
Sackville Street, P.O. Box 88, Manchester M60 1QD, U.K.\\
$^{12}$Department of Physics, Macquarie University, NSW 2109, Australia\\
$^{13}$Anglo-Australian Observatory, PO Box 296, Epping NSW 1710, Australia\\
$^{14}$Departamento de Fisica Teorica y del Cosmos,
Facultad de Ciencias, Avd. Fuentenueva S/N, Granada, 18071, Spain\\
\\
}

\date{received,  accepted}

\maketitle

\begin{abstract}
The INT Photometric H$\alpha$ Survey of the Northern Galactic Plane (IPHAS) is 
a 1800 square degrees CCD survey of the northern Milky Way spanning the
latitude range $-5^{\rm o} < b < +5^{\rm o}$ and reaching down to 
$r' \simeq 20$ (10$\sigma$).  It may increase the number of known northern 
emission line sources by an order of magnitude.  Representative observations 
and an assessment of point-source data from IPHAS, now underway, are presented.
The data obtained are Wide Field Camera images in H$\alpha$ alpha narrow-band, 
and Sloan $r'$ and $i'$ broad-band, filters.  We simulate IPHAS 
$(r' - H\alpha, r' - i')$ point-source colours using a spectrophotometric 
library of stellar spectra and available filter transmission profiles: this 
gives expected colours for (i) solar-metallicity stars, without H$\alpha$ 
emission, and (ii) emission line stars.  Comparisons with Aquila field 
observations show that  simulated normal star colours reproduce the data well 
for spectral types earlier than M. Spectroscopic follow-up of a Cepheus field 
confirms that sources lying above the main stellar locus in the 
$(r' - H\alpha, r' - i')$ plane are emission line objects, with very few 
failures.  Examples of H$\alpha$ deficit objects -- a white dwarf and a carbon 
star -- are shown to be readily distinguished by their IPHAS colours.  The 
role IPHAS can play in studies of nebulae is discussed briefly, and 
illustrated by a continuum-subtracted mosaic image of the SNR, Shajn~147. The 
final catalogue of IPHAS point sources will contain photometry on about 80 
million objects. (abridged for astro-ph))  
\end{abstract}


\section{Introduction}

The astronomical significance of $H\alpha$ spectral line emission is that it 
both traces diffuse ionized nebulae and is commonly prominent in the spectra 
of pre- and post-main-sequence stars and binaries.  Since these are objects
in relatively short-lived phases of evolution, they are a 
minority in a mature galaxy like our own.  Their scarcity has in turn acted as 
a brake on our understanding of these crucial evolutionary stages that in 
youth help shape the growth of planetary systems, and in old age determine 
stellar 
end states and the recycling of energy and chemically-enriched matter back 
into the galactic environment.  The major groups of emission line stars 
include all evolved massive stars (supergiants, luminous blue variables, 
Wolf-Rayet stars, various types of Be star), post-AGB stars, pre-main-sequence 
stars at all masses, active stars and interacting binaries. This last group 
most likely harbours SN~Ia progenitors within it, in guises that are still
subject to considerable debate (Hillebrandt \& Niemeyer 2000; Uenishi, Nomoto
\& Hachisu 2003).

     Existing catalogues of emission line objects contain anything from a 
few, to a few hundred, sources.  Within the least populous object 
classes (e.g. the luminous blue variables and supersoft X-ray binaries, with 
just a few of each known in the Galaxy) there can be a confusing m\^el\'ee of 
`special cases' that inhibit confident identification of essential and general 
behaviours.  In effect, stellar evolutionary studies have been bedevilled by 
small number statistics and a lack of good demographics.  The remedy for this 
problem is to exploit the technical developments of recent years that have 
boosted both the efficiency with which large scale astronomical surveys can be 
performed, and the quality that can be achieved.  In particular, large area 
CCD-mosaic detectors offering good spatial resolution have now completely 
supplanted the photographic techniques of the last century.  In this paper, 
we describe the Isaac Newton Telescope (INT) Photometric H$\alpha$ Survey of 
the Northern Galactic Plane (IPHAS), a programme that began taking data with 
the INT Wide Field Camera in the second half of 2003.  

     The goal of IPHAS is to survey the entire northern Galactic Plane in the 
latitude range $-5^{\rm o} < b < +5^{\rm o}$ -- a sky area of 1800 sq.deg.
The choice of latitude range was tensioned between the rising total telescope 
time requirement and the expected fall off in discoveries to be made with 
increasing latitude (see below).  The 10-degree wide strip requires in the 
region of 22 weeks clear time, and the hope is to complete the observations
before the end of 2006.  The data obtained will be mined both for 
spatially-resolved nebulae and for unresolved emission line stars.  For point 
sources, the magnitude range will be $13 \lesssim r' \lesssim 20$.  Here 
we will focus on presenting the basic features of the survey, together with 
the extraction of point source data and the analysis of photometric colour 
information.  The different technical issues relating to the 
identification and measurement of resolved H$\alpha$-emitting nebulae will be 
presented in a later paper.  For now, we just point to the opportunity that 
IPHAS presents both for making new discoveries and for high quality H$\alpha$ 
emission mapping on large angular scales.

     To place this new northern hemisphere survey in context, it is 
appropriate to review the scale and character of the emission line star 
population that previous Galactic H$\alpha$ surveys have revealed. 
Kohoutek \& Wehmeyer (1999, hereafter KW99) have added their own discoveries 
within the latitude range $-10^{\rm o} < b < +10^{\rm o}$ (1979 objects; data 
obtained in the years 1964-1970) to those of a wide range of independent 
searches: these go back as far as the original work of Merrill \& Burwell that 
resulted in the Mount Wilson Catalogue (MWC, see Merrill \& Burwell 1933).  
The total number of KW99 objects is 4174.  In many cases the source 
observations are spectra obtained using objective prism facilities.  For 
sources in the northern hemisphere, this compilation supercedes that due to 
Wackerling (1970). Three-quarters of the stars listed in KW99 are assigned a 
photovisual magnitude, $m_{\rm pv} < 13$, and it is surmised that this is 
roughly the catalogue's completeness limit.  They also note that over 80 
percent of all the objects they list in the $-10^{\rm o} < b < +10^{\rm o}$ 
band fall within the narrower $-5^{\rm o} < b < +5^{\rm o}$ band.  At the 
fainter magnitudes we are exploring, we might expect this concentration toward 
the Galactic Equator to become even more pronounced.  Naive extrapolation of 
the bright-end ($m_{\rm pv} < 13$) magnitude distribution of the KW99 emission 
line stars to span $13 < m_{\rm pv} < 20$ would suggest that our survey should 
uncover 8000--10000 new objects.

     This is probably an underestimate for a number of reasons. First,
we can check this extrapolation of KW99 against the same quantity
derived from the Stephenson \& Sanduleak (1971, hereafter SS71) southern 
Galactic Plane survey.  The SS71 completeness limit is shallower at 
$m_{\rm pg} \sim 11$.  We find, even on excluding the Galactic Bulge region 
located exclusively in the southern sky, that the prediction rises to 
$\sim$40000. This dramatic difference can have a number of origins -- 
beginning with simple differences in the Galactic stellar populations 
accessible from the northern and southern hemispheres, and ending with issues 
of experimental technique.  Nevertheless a parallel between the KW99 and SS71 
catalogues is that the bright magnitudes sampled strongly favour early-type, 
intrinsically luminous stars (such objects account for three-quarters of the 
KW99 catalogue).  On going to much fainter magnitudes the sampled emission 
line star population is likely to broaden in character as intrinsically 
fainter object types (e.g. young and active stars, interacting binaries) 
become included.  

     An immediate precursor to IPHAS and, indeed, a prompt for the need for
a northern survey, is the AAO/UKST narrow-band H$\alpha$ Survey of the 
Southern Galactic Plane and Magellanic Clouds.  This was the last photographic 
sky survey carried out on the UK Schmidt Telescope (UKST). It was completed 
in 2003 and is now available as digital survey data derived from SuperCOSMOS 
scans of the original survey films (the SHS database, located at 
http://www-wfau.roe.ac.uk/sss/halpha/).  A description of this survey is 
presented by Parker et al (2005): important points to note are its high 
spatial resolution ($\sim 1$~arcsec) and its areal completeness -- the 
entire southern Galactic Plane was imaged within the latitude range 
$-10\deg < b < +10\deg $.  Each of the 233 Galactic Plane fields observed had 
an effective dimension projected on the sky of $4\deg \times 4\deg$.  
The southern survey has provided the source material for a variety of 
continuing research projects (see e.g. Morgan, Parker \& Russeil 2001; 
Parker \& Morgan 2003; Drew et al 2004).  For the detection of point sources, 
IPHAS betters both the SHS sensitivity and spatial resolution, and offers the 
advantage of CCD dynamic range and linearity.  The sensitivity of the two 
surveys to spatially-resolved H$\alpha$ emission is comparable.

\begin{table*}
\caption{Co-ordinates of centres, estimates of maximum Galactic reddenings,
observation dates and seeing for the IPHAS fields discussed in this paper.
Note that the offset partner field (nnnno to field nnnn) is offset by 5 arcmin
N and 5 arcmin E.  The reddening estimates are derived from the Schlegel et al 
(1998) reddening maps.} 
\label{field_list}
\begin{tabular}{lccrrcccl}
\hline
IPHAS & \multicolumn{4}{c}{Field Centre Coordinates} & maximum & 
  observation & mean & \\
field & RA & Dec & $\ell$ & $b$ & reddening & date & seeing & \\
number & (2000) & (2000) & ($\deg $) & ($\deg $) & [$E(B-V)$] & 
 (dd/mm/yyyy) & (arcsec) & \\   
\hline  
2540 & 05 33 49 & $+$25 15 & 181.73 & $-$4.18 & 1.1 & 05/11/2003 & 0.8 &
Section 2, 5 \\
4090 & 18 32 03 & $+$00 41 & 31.33 & $+$4.62 & 2.3 & 09/06/2004 & 1.1 &
Section 3 \\
4095 & 18 33 33 & $+$01 47 & 32.48 & $+$4.79 & 1.6 & 09/06/2004 & 1.0 &
Section 3, 4 \\
4199 & 18 47 27 & $+$01 58 & 34.23 & $+$1.78 & 3.0 & 12/06/2004 & 0.9 &
Section 3, 4 \\
6985 & 22 14 56 & $+$61 11 & 105.18 & $+$3.83 & 2.3 & 05/11/2003 & 0.8 & 
Section 6 \\
6993 & 22 15 56 & $+$61 44 & 105.59 & $+$4.21 & 2.5 & 05/11/2003 & 0.8 & 
Section 6 \\
7012 & 22 18 36 & $+$61 22 & 105.65 & $+$3.73 & 1.8 & 03/11/2003 & 1.2 &
Section 6 \\
7019 & 22 19 39	& $+$61 55 & 106.06 & $+$4.12 & 2.4& 03/11/2003 & 1.2 &
Section 6 \\
\hline
\end{tabular}
\end{table*}

     We begin the description of IPHAS in the next section with
a presentation of the filters used, and our observing and data reduction
techniques.  Following this, in section 3, we discuss the use of the 
H$\alpha$,$r'$,$i'$ filter photometry in the diagnostic 
$(r'-H\alpha)$ versus $(r'-i')$ diagrams that can be constructed from the
survey data for point sources.  Specifically, we introduce simulated 
colour-colour tracks for both normal solar-metallicity stars occupying the 
main stellar locus and for emission line objects.  We then provide examples of 
$(r' - H\alpha, r' - i')$ diagrams in three contrasting northern 
Galactic plane locations (sections 4 to 6).  The fields discussed are 
identified in Table~\ref{field_list}.  In section 4, we illustrate
the application of the simulated tracks for normal stars with reference
to fields in Aquila; in section 5 we perform a consistency check of IPHAS 
photometry of a Taurus field obtained on a photometric night; and in 
section 6 we present some follow-up spectroscopy relating to a field in 
Cepheus that illustrates the high success rate achieved in the confirmation 
of candidate emission line objects.  In Section 7, we outline the application
of IPHAS to imaging spatially-resolved nebulae, and illustrate this with
the beautiful example of the supernova remnant, S~147.  The paper ends with a 
summarising discussion (Section 8).

\section{Survey observations and data extraction}

\subsection{IPHAS observations}

     The Wide Field Camera, mounted on the 2.5-metre Isaac Newton Telescope, 
is an imager comprising 4 AR-coated, thinned 4K $\times$ 2K EEV CCDs arranged 
in an L shape, capturing data from an on-sky area of approximately 0.3 of a 
square degree.  With a pixel dimension of 13.5~$\mu$m, corresponding on-sky to 
0.333$\times$0.333 arcsec$^2$, the instrument is appropriately configured to 
fully exploit the high quality sub-arcsecond seeing frequently encountered at 
the Roque de los Muchachos Observatory in La Palma.  Adequately sampled 
$\sim$1 arcsec resolution is particularly useful given that lower-reddening 
Galactic Plane star fields, observed down to $\sim$20th magnitude, are at 
times very crowded.

     Not accounting for the geometric consequences of the L-shaped detector 
arrangement, the total number of pointings required to span the 
10$\times$180~deg$^2$ survey area would be 6000.  On accounting for the 
detector outline and requiring a little overlap between pointings, we have 
chosen to fix the number of field centres at a total of 7635.  Furthermore,
each pointing is paired with a second pointing at an offset of 5 arcmin W and 
5 arcmin S, such that the number of quality-controlled sets of exposures
expected to be compiled into the final survey database is 15270.  
Stars falling into a gap between the mosaiced CCDs in one exposure are 
captured in a partner exposure set.  Nevertheless the great majority of 
Galactic Plane sources will be imaged at least twice. The pairing and 
offsetting, together with the chosen tessellation, comes very close to 
complete coverage of the northern Plane ($>$ 99 \%). 

     Since H$\alpha$ falls in the red part of the spectrum, IPHAS was 
conceived of as a large-scale programme that could readily make use of 
less heavily subscribed bright and grey nights. To ensure this scheduling 
flexibility, whilst obtaining the associated continuum-band observations 
required for establishing unambiguous H$\alpha$ excesses, it was decided
to restrict our broadband choices to red or longer wavelengths.  This also
has the effect of increasing the penetration of the survey for a given
exposure time since these longer wavelengths are also less subject to 
Galactic dust obscuration than UBV bands. This stands in contrast to the 
somewhat bluer emphases of the older H$\alpha$ catalogues (e.g. SS71,
where the objective prism data spanned $3300 < \lambda$(\AA )$ < 6800$).  
The particular choice we made was to obtain the H$\alpha$ exposures alongside 
Sloan $r'$ and $i'$ filter observations.  

\begin{figure}
\begin{picture}(0,190)
\put(0,0)
{\includegraphics{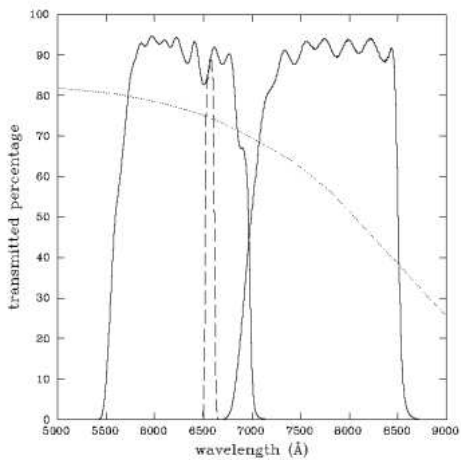}}
\end{picture}
\caption{The transmission profiles of the $H\alpha$, Sloan $r'$ and $i'$ 
filters used in all IPHAS observations.  The $r'$ and $i'$ filter profiles are 
plotted as solid lines, whilst the $H\alpha$ profile is shown dashed.  The 
dotted line is the mean WFC CCD response at the cooled working temperature.
These profiles are available in numerical form from the INT/WFC web page,
http://www.ing.iac.es/~quality/filter/filt4.html}  
\label{filters}
\end{figure}

     All three filter profiles are plotted as Fig.~\ref{filters}.  The 
Sloan filters have been preferred over Harris alternatives because of their 
squarer transmission profiles.  The Sloan $r'$ filter is the most 
blue-sensitive of the three (central wavelength 6240~\AA ), with the H$\alpha$ 
filter positioned toward the red end of its bandpass (central wavelength 
6568~\AA ).  With a full-width at half-maximum (FWHM) transmission of 95~\AA ,
the H$\alpha$ filter is more than broad enough to capture all likely Doppler 
shifts due to Galactic motions of up to a few hundred km~s$^{-1}$ or 
$\sim$10~\AA , as well as blueshifts of up to an additional $\sim$10~\AA\ due
to the converging beam of the INT/WFC. The central wavelength of the Sloan 
$i'$ filter is 7743~\AA .  We have added two broad band filters to this survey 
in order to give a continuum-dominated colour with which the $(r'-H\alpha)$ 
excess measurement can be compared.  It has been shown in the past (see e.g. 
Robertson \& Jordan 1989) that this is important for distinguishing between a 
genuine H$\alpha$ emission excess and a molecular band dominated late type 
stellar spectrum -- in such cases large $(r'-H\alpha)$ `colour' will correlate 
with relatively extreme $(r'-i')$.  In truth, the diagnostic value of this 
strategy is wider than this, as shall become apparent in Section 3. 

The exposure times in the three filters were set at 120~sec ($H\alpha$) and 
10~sec ($r'$ and $i'$) for the first season's observing in 2003.  Evaluation 
of these data, once extracted, led us to increase the $r'$ band exposure to 
30~secs, from the start of the 2004 observing season, to compensate better for 
their typically higher moonlit background.  This adjustment also 
acknowledges the pivotal role the $r'$ band exposures must play in the 
survey's exploitation -- it is important that errors in this band, appearing 
in both the H$\alpha$ excess and the broadband colour measurement, are 
minimised.  

     For the purpose of photometric calibration, each night's observations
includes standard fields, obtained in twilight and at intervals of 
approximately 2 hours through the night.  The standards are chosen from
a list including the Landolt equatorial fields (Landolt 1992), Sloan
(Smith et al 2002) and Stetson standards (at the Canadian Astronomy Data 
Centre, http://cadcwww.dao.nrc.ca/standards/).  Nightly observations
are also acquired of spectrophotometric standards with a view to assisting
the final calibration of the narrow-band H$\alpha$ data.   A programme of 
supporting spectrophotometric observations is planned with a view to placing 
the H$\alpha$ calibration on the desired firm footing in the longer term. 

\subsection{Data processing}

   Processing of IPHAS INT WFC data generally follows the pipeline procedure 
devised by Irwin and Lewis (2001) for dealing with optical mosaic camera 
data.  The two-dimensional instrumental signature removal includes provision 
for: non-linearity correction at the detector level; bias and overscan 
correction prior to trimming to the active detector areas; flatfielding;
and fringe removal in the $i'$ passband.  

Flatfielding in all bands is accomplished by stacking suitable twilight 
flatfield exposures taken over the course of each typically one-week observing 
run to create master calibration flats.  These have been found to be stable 
on this timescale provided no filter changes, or other instrumental setup 
changes, occur in the middle of the run.  The gain differences between each 
detector in each passband are removed by normalising a robust measure of the 
average sky level for each detector to a common system (in this case the sky 
level on CCD no. 1).  The flatfielded $i'$ data, even for the short exposures 
(10~s) used here, show measurable fringing.   At the same time, the 
data taken for the IPHAS project are not themselves suitable (short exposures 
in crowded Galactic Plane regions) to construct good quality fringe maps for 
correcting this problem.  Since the fringing in the $i'$ band is relatively 
stable with time, we make use of a library of $i'$ band fringe maps taken from 
other observing runs using the INT WFC.  These have been found to reduce the 
level of fringing to an acceptable level when used with the defringing 
algorithm in the pipeline.

Each master flat, in conjunction with a previously defined bad column list, 
is also used to construct confidence maps for each passband.  These are used 
during the catalogue generation to flag less reliable pixels in each image by 
providing a measure of the inverse variance weight for each pixel e.g. bad 
pixels have zero weight, heavily vignetted regions have low weight,
poor DQE pixels have lower weight, and so on.  These confidence measures are 
used directly to weight the image detection part of the catalogue generation
algorithm and help avoid generating excessive numbers of spurious images 
around defects and other excessively noisy regions.

Catalogue generation follows the precepts outlined by Irwin (1985, 1997)
and includes the facility to: automatically track any background variations
on scales of typically 20--30 arcsec; detect and deblend images or groups
of images; and parameterise the detected images to give various (soft-edged)
aperture fluxes, position and shape measures.  The generated catalogues 
start with an approximate World Coordinate System (WCS) defined by the 
known telescope and camera properties (eg. WCS distortion model) and are 
then progressively refined using all-sky astrometric catalogues (eg. USNO, 
APM, 2MASS) to give internal precision generally better than 0.1 arcsec and 
global external precision of ~0.25 arcsec with respect to USNO and APM, and 
~0.1 arcsec with respect to 2MASS.  These latter numbers are solely dependent 
on the accuracy of the astrometric catalogues used in the refinement.

All catalogues for all CCDs for each pointing are then processed
using the image shape parameters for morphological classification in the
main categories: stellar; non-stellar; noise-like.  A sampled curve-of-growth 
for each detected object is derived from a series of aperture flux measures as 
a function of radius.  The classification is then based on comparing the 
curve-of-growth of the flux for each detected object with the well-defined 
curve-of-growth for the general stellar locus.  This latter is a direct 
measure of the integral of the point spread function (PSF) out to various 
radii and is independent of magnitude, {\em if} the data are properly 
linearised, and if saturated images are excluded.  The average stellar locus 
on each detector is clearly defined and is used as the basis for a null 
hypothesis stellar test for use in classification.  The curve-of-growth for 
stellar images is also used to automatically estimate frame-based aperture 
corrections for conversion to total flux. 
\footnote{We note that in regions of intense nebular emission with 
increasingly 
short spatial scale variations of the "background", automatic detection,
parameterisation and classification of objects becomes progressively
more unreliable.  In such regions continuum subtraction via difference
imaging will yield better results.}

Any photometric standards observed during the run (mainly Landolt 1992
and spectrophotmetric standards) are automatically located in a standards
database and used to estimate the zero-point in each passband for every
pointing containing any of these standards.  The trend in the derived
zero-points is then used to assign a photometric quality index for
each night and also as a first pass estimate for the magnitude calibration
for all the observations.  The H$\alpha$ filter is treated as equivalent to
a standard Johnson-Cousins R-band filter to obtain a Vega-like magnitude 
which is used as an initial calibration (to be refined later, as mentioned 
above).

\begin{figure}
\begin{picture}(0,270)
\put(0,0)
{\includegraphics{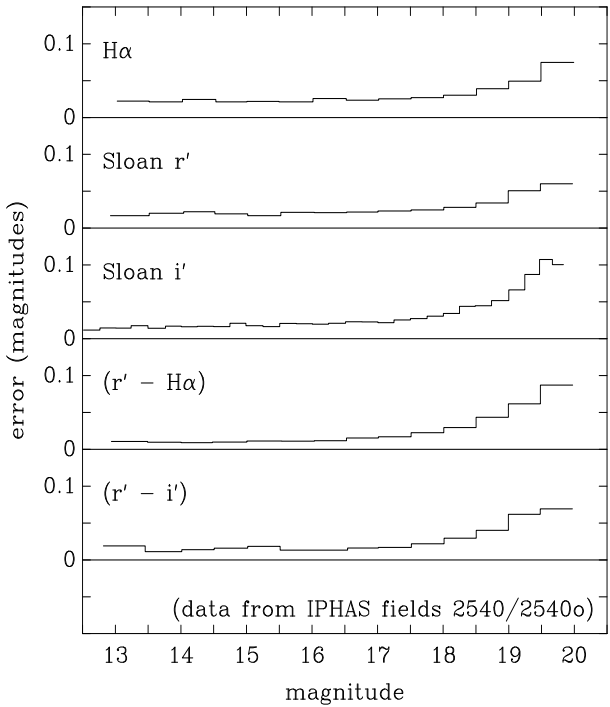}}
\end{picture}
\caption{Magnitude and colour errors in IPHAS data as a function of magnitude.
The quantity shown, for each filter in the top 3 panels, is the rms deviation 
per magnitude bin, between the measured magnitudes for each of IPHAS 
fields 2540 and 2540o.  The lower two panels show the colour rms deviations.
The bin size is 0.5 magnitudes for all but the $i'$ data, where a binning of 
0.25 magnitudes was more convenient.  See further discussion of fields 
2540/2540o in section 5.  These data were obtained on a photometric night in 
November 2003.  The $r'$ exposure times were 10~sec in this case.} 
\label{errors}
\end{figure}

Various quality control plots are generated by the pipeline and these
are used to monitor characteristics, such as:- the seeing; the average
stellar image ellipticity (to measure trailing); the sky brightness
and sky noise; the size of aperture correction for use with the ``optimal''
aperture flux estimates (here ``optimal'' refers to the well-known property
that soft-edged apertures of roughly the average seeing radius provide
close to profile fit accuracy eg. Naylor 1998).   The ``optimal'' catalogue 
fluxes for the $r'$, $i'$ and H$\alpha$ filters for each 
field are then combined to produce a single matched merged catalogue from
which diagnostic colour-magnitude diagrams and two-colour diagrams 
may be produced.  These merged catalogues -- the fundamental IPHAS product 
-- contain flux, classification and match position error for each
object in each passband.  

The IAU-registered naming convention for all point sources derived from these 
catalogues is IPHAS JHHMMSS.ss$+$DDMMSS.s -- thereby encoding the 2000 object 
co-ordinates into the name.  

To give an impression of the internal magnitude errors in the catalogued 
magnitudes and derived colours we plot, in Fig.~\ref{errors}, the rms 
deviation between the magnitudes measured in each filter, and the associated 
colours, for point sources common to two overlapping exposure sets (fields
2540 and 2540o, discussed again in Section 5).  These were obtained on a 
photometric night in November 2003 as the moon was setting.  Calculated 
empirically as $\sqrt{<(m_{\rm {2540}} - m_{\rm {2540o}})^2>}$ or its colour 
equivalent, over a range in mean magnitude $\Delta m = 0.5$ or 
$\Delta m = 0.25$, the error is corrected back to a representative single 
field measurement error by dividing by $\sqrt{2}$.  The bright-end errors in 
the magnitudes themselves, in plots such as these, are typically dominated by 
calibration offsets of a few hundredths that will be removed when a final 
uniform survey calibration is devised.   In the case of fields 2540 and 2540o 
the offsets were all small (less than $\sim 0.01$). From 2004 on, when the 
$r'$ exposures were increased to 30~sec, the faint-end $r'$ errors drop to 
around 60\% of those for 2003 (for the same sky conditions). This carries 
through to the colour errors falling to 80\% or less of their 2003 levels. 
Altogether, this significantly raises the fraction of catalogued objects that 
will meet the quality target of $\Delta r' \leq 0.1$ for $r' \leq 20$.

To date roughly 3 Tbytes of raw data from the first two seasons of
IPHAS observing have been processed this way.  This corresponds to well over 
100,000 4kx2k CCD images and over 40 million objects have been catalogued.
All of this processed data is also available at the individual frame and 
catalogue level via a PostgreSQL database interface which allows users to:
postage stamp browse for candidate verification; construct image catalogue 
overlays, including on-the-fly matching with other catalogues such as the 
2MASS point source catalogue; perform on-demand continuum image subtraction 
and mosaicing of larger areas; access all the quality control information; 
and more (see Irwin et al 2005).  The database interface is available
on the Cambridge Astronomical Survey Unit (CASU) website, at
http://apm2.ast.cam.ac.uk/cgi-bin/wfs/dqc.cgi.  Co-ordinates of the centres 
of the observed IPHAS fields are obtainable there.
\footnote{These are identified via object names taking the form 
intphas\_nnnn$\ast$, where nnnn is a 4-digit number up to 7635, and 
$\ast$ is the wild card for further characters identifying exposure type.}
      
\section{Simulation of the IPHAS colour-colour plane}
\label{simul}

   The three bandpasses of the survey provide the basis for the construction
of a number of magnitude-colour diagrams and a colour-colour diagram to 
describe any chosen region in the northern Galactic Plane.  Using just the 
two $r'$, $i'$, broad bandpasses, one may derive colour-magnitude diagrams 
that can in principle reveal different sequences at different reddenings that 
may be present in the field under investigation. 

   Full exploitation of IPHAS hinges on the colour-colour plane involving
all three bands.  The combination of magnitudes we use is $(r' - i')$ as 
abscissa and $(r' - H\alpha)$ as ordinate, so that objects with H$\alpha$
band excesses appear higher within the diagram, while intrinsically redder or 
more highly reddened objects are over to the right.  The most straightforward 
use that can be made of such diagrams is to pick out for spectroscopic 
follow-up those objects whose $(r' - H\alpha)$ colour places them clearly 
above the main locus of non-emission line objects.  Additional information
contained within the colour-colour diagrams can lead to identification
of more subtle candidate emission line stars and also to a characterisation 
of the stellar populations distributed along the line of sight.  In this 
second sense, IPHAS can also be seen as providing a far-red map of stellar 
populations in the northern Galactic Plane.

\begin{figure}
\begin{picture}(0,210)
\put(0,0)
{\includegraphics{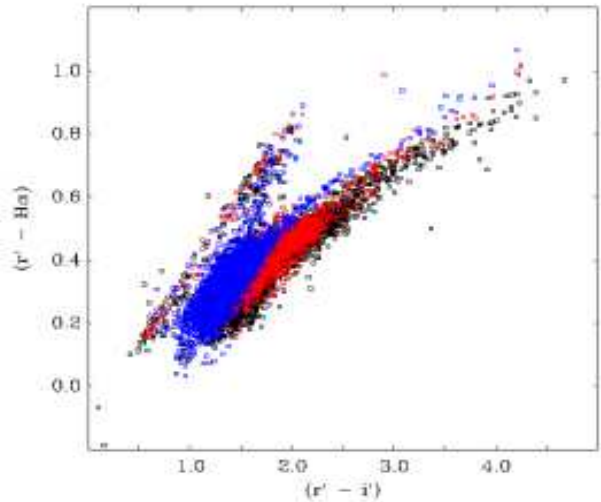}}
\end{picture}
\caption{A plot of $(r' - H\alpha)$ versus $(r' - i')$ for 13,818
point sources extracted from the overlap area of the paired exposures
for IPHAS fields 4199 (black points), 4090 (red points) and 4095 (blue
points), in Aquila.  The sources lie in the magnitude range 
$13 < r' < 20$.  The colours for each object have been formed independently
from each set of $H\alpha$, $r'$ and $i'$ exposures, and then averaged.
The median error in either colour is $\sim$ 0.03 at $r'$ close to 19,
and only occasionally exceeds 0.06 at the $r' = 20$ limit of the plot.
Note the very well defined upper edge to the upper sequence -- this is
where unreddened main sequence stars are located.  The few stars above it are
candidate emission line objects.}  
\label{aquila_cc}
\end{figure}

   Fig.~\ref{aquila_cc} is a composite $(r'- H\alpha, r' - i')$ 
plot derived from data obtained in three paired IPHAS fields (fields 
4090/4090o, 4095/4095o and 4199/4199o: see Table~\ref{field_list}).  In each 
case, catalogues of sources classified by the CASU pipeline as either 
`definitely stellar' or `probably stellar' were extracted from within a 
30 x 30 arcmin$^2$ box spanning most of the overlap region between the two 
pointings.  For each extracted object, the datum plotted is the mean of the 
colours derived independently from each of the two exposures making up the 
field pair.  The data shown are limited to the magnitude range $13 < r' < 20$, 
where the error in either colour is kept to less than $\sim$0.05 magnitudes.  
These are representative of the better data in the IPHAS database in that they 
were obtained on photometric nights in June 2004 at times of $\sim$1~arcsec 
seeing and low sky background.  

    All three fields are located in the Aquila Rift region, and 
sample sightlines that pass through the outer parts of the molecular cloud.  
Dame \& Thaddeus (1985) noted that this is a nearby ($\sim$200 pc) and not  
particularly opaque cloud system, presenting around 2 magnitudes of visual 
extinction only.  This is a modest addition to the reddening through the 
remaining Galaxy beyond -- the reddening data of Schlegel, Finkbeiner \& 
Davis (1998) indicate maximum visual extinctions, $A_V$ ranging from $\sim 5$, 
in 4095, up to $\sim 10$, in 4199.  The nearby rift cloud is responsible for 
the lightly-populated gap, seen in Fig.~\ref{aquila_cc}, between the upper 
sequence and the lower, but much more densely populated strip.  The existence 
of this separation allows a clear demonstration of how well 
theoretically-synthesised tracks compare with and make sense of the 
photometry.  

   To achieve an understanding of the behaviours seen in the colour-colour
domain, we have constructed two types of synthetic tracks: the first type
concerns the properties of normal stars without H$\alpha$ emission, while
the second explores the impact of adding narrow H$\alpha$ emission to
generic stellar spectral energy distributions (SEDs).  We present these 
tracks below, using the Aquila fields to illustrate the former in section 4.

\subsection{The IPHAS colours of normal stars}
\label{normal}

For simulating the $(r' - H\alpha)$ and $(r' - i')$ colours of 
normal stars, we have used the library of stellar spectral energy 
distributions (SEDs) due to Pickles (1998, hereafter P98).  
At a final binning of 5~\AA\, the spectra in this library are well enough 
sampled that we may use them to compute narrow-band $H\alpha$ relative 
magnitudes with confidence, alongside the analogous broadband $r'$ and $i'$ 
quantities.  The required numerical filter transmission profiles, shown in 
Fig.~\ref{filters}, are available via the ING WFC web pages 
(http://www.ing.iac.es/Astronomy/instruments/wfc/), as is a mean 
Wide Field Camera CCD response curve.  To ensure compliance with the 
Vega-based zero magnitude scale, we have defined synthetic colour as follows:
\begin{equation}
  (r' - i') = -2.5 \log (\frac{\Sigma T_r' F_{\lambda} \Delta\lambda}{\Sigma T_r'
          F_{\lambda ,V} \Delta\lambda}) + 2.5 \log (
          \frac{\Sigma T_i'F_{\lambda} \Delta\lambda}{\Sigma T_i'
          F_{\lambda ,V} \Delta\lambda})
\label{eqn_colour}
\end{equation}
where $T_r'$ and $T_i'$ are the $r'$ and $i'$ numerical transmission profiles,
after multiplying by the mean WFC CCD response curve, and rebinning 
to match the P98 spectral library sampling.  The SED for Vega, 
$F_{\lambda ,V}$, is the appropriately resampled version of that due to Hayes 
(1985).  The $(r' - H\alpha)$ colour is evaluated in the same way, after 
substituting the $H\alpha$ numerical profile in place of the $i'$ profile.
Since Vega is an A0V star, its SED at H$\alpha$ incorporates a strong 
absorption line feature.  Currently, because the CASU pipeline uses 
broad-band standard fields to calibrate measured source magnitudes, there
is an offset in $(r' - H\alpha)$ colour between the catalogue data and
our simulations.  The dominant spectral type in the standard fields will
be appreciably later than Vega's A0, with the consequence that the 
standard-star SEDs will both be redder and less eroded by H$\alpha$ line 
absorption.  On this basis one would expect, and we do find, that zero 
$(r' - H\alpha)$ for unreddened main sequence stars corresponds to 
$(r' - i') \sim 0.3$ (late F), rather than to $(r' - i') = 0$ (Vega, A0V), in 
plots of IPHAS data obtained in photometric conditions.  Hence, on comparing 
simulated tracks with observation it is necessary to correct for this.  The 
best way to do this is to assume that only a shift in $(r' - H\alpha )$ is 
required, given that the calibration of the broadband-only $(r' - i')$ colour 
should be secure enough (see Section~5 and the discussion of 
Fig.~\ref{2540_cc}).  The shift that needs to be applied to theoretical 
$(r' - H\alpha )$ values to match them to observation is then always 
downwards, varying in amount between about $-$0.10 and $-$0.25.

\begin{figure*}
\begin{picture}(0,360)
\put(0,0)
{\includegraphics{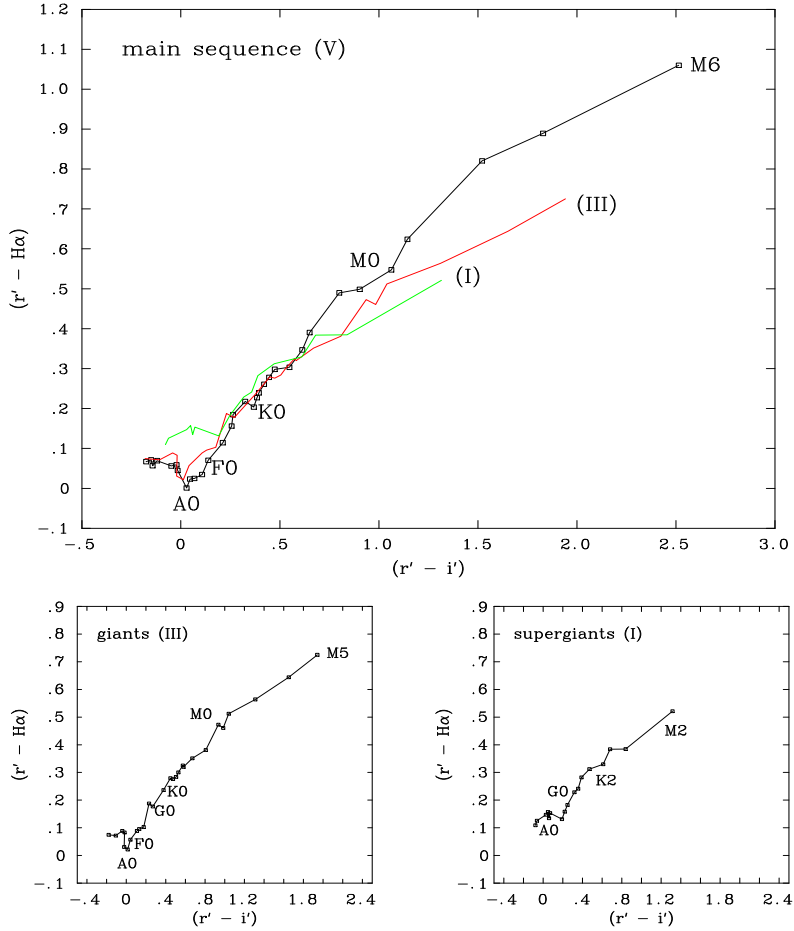}}
\end{picture}
\caption{The expected positions of unreddened main sequence (top), 
giant (bottom left) and supergiant stars (bottom right) in the 
$(r' - H\alpha, r' - i')$ plane according to spectral type.  In the top panel 
the giant and supergiant tracks (respectively red and green) are superimposed 
to show their positioning relative to the main sequence.} 
\label{tracks_0}
\end{figure*}

\begin{figure*}
\begin{picture}(0,190)
\put(0,0)
{\includegraphics{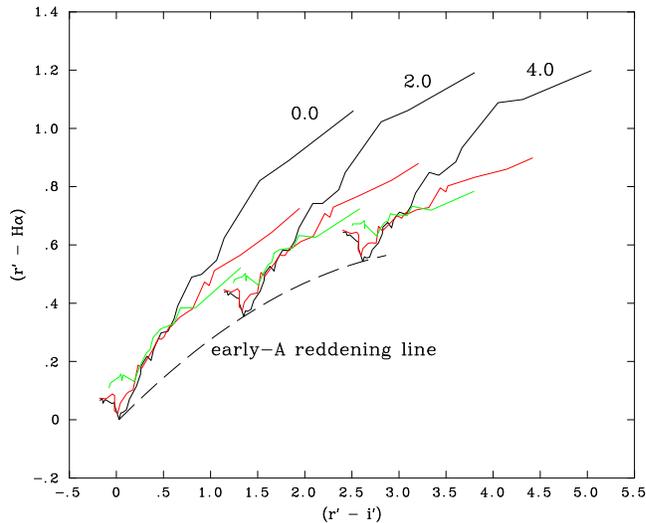}}
\end{picture}
\caption{The effect of interstellar extinction, calculated according to an
$R = 3.1$ Galactic law, on synthetic stellar tracks in the 
$(r' - H\alpha, r' - i')$ plane.  As in Fig.~\ref{tracks_0}, the main
sequences are drawn in black, the giant sequences in red, and supergiants
in green.  The three sets shown apply to $E_{B-V} = 0.0$, $2.0$ and $4.0$,
as labelled.  The dashed line shown is the reddening locus for A0V stars.
This defines a notional minimum line for all non-degenerate stars.  It is 
named the ``early-A reddening line'' for reasons that become clear in 
Section 5.}
\label{tracks_red}
\end{figure*}

    We have simulated tracks for main sequence stars (luminosity class V),
giant stars (class III) and supergiants (class I) using, for simplicity only, 
solar metallicity P98 spectra.  Each sequence has been calculated for a 
range of reddenings using an $R=3.1$ optical-IR extinction law in the form
given by Howarth (1983).  The colours derived are given for $E_{B-V} = 0$, 1,
2, 3 and 4 in Tables~\ref{cc_tracks_1} and \ref{cc_tracks_2}.  They are 
specified at this level of detail because there is no single reddening 
vector that translates all of a sequence directly onto its reddened 
counterpart.  The unreddened dwarf,
giant and supergiant sequences are compared in Fig.~\ref{tracks_0}, where 
it can be seen that there is a gradual decrease in track gradient with 
increasing luminosity class.  Nevertheless between mid-F and late-K spectral 
types there is minimal distinction between the luminosity classes -- this is 
the regime where H$\alpha$ absorption is weak and there is not yet any marked 
development of the molecular band structure that typifies M-type stars.

\begin{table*}
\caption{Synthetic tracks in the $(r' - H\alpha, r' - i')$ plane, for 
main-sequence dwarfs and giants, calculated for a range of reddenings.}
\label{cc_tracks_1}
\begin{tabular}{lrlllllllll}
\hline
Spectral & \multicolumn{4}{l}{Reddening:-} & & & & & & \\
Type     & \multicolumn{2}{l}{$E_{B-V} = 0.0$} & \multicolumn{2}{l}{$E_{B-V} = 1.0$} &
           \multicolumn{2}{l}{$E_{B-V} = 2.0$} & \multicolumn{2}{l}{$E_{B-V} = 3.0$} &
           \multicolumn{2}{l}{$E_{B-V} = 4.0$} \\
 & $(r'-i')$ & $(r'-H\alpha)$ & $(r'-i')$ & $(r'-H\alpha)$ & $(r'-i')$ 
 & $(r'-H\alpha)$ & $(r'-i')$ & $(r'-H\alpha)$ & $(r'-i')$ & $(r'-H\alpha)$ \\
\hline
 & & & & & & & & & & \\ 
O5V  &  -0.176 &   0.067 &   0.493 &   0.274 &   1.146 &   0.438 &   1.784 &   0.560 &   2.408 &   0.643 \\
O9V  &  -0.151 &   0.072 &   0.519 &   0.278 &   1.173 &   0.441 &   1.811 &   0.563 &   2.436 &   0.645 \\
B0V  &  -0.143 &   0.057 &   0.529 &   0.264 &   1.184 &   0.427 &   1.824 &   0.549 &   2.451 &   0.632 \\
B1V  &  -0.120 &   0.069 &   0.552 &   0.273 &   1.209 &   0.435 &   1.850 &   0.554 &   2.478 &   0.635 \\
B3V  &  -0.049 &   0.056 &   0.623 &   0.258 &   1.279 &   0.418 &   1.919 &   0.535 &   2.546 &   0.614 \\
B8V  &  -0.022 &   0.059 &   0.649 &   0.259 &   1.304 &   0.416 &   1.944 &   0.532 &   2.572 &   0.609 \\
B9V  &  -0.016 &   0.045 &   0.652 &   0.243 &   1.305 &   0.399 &   1.943 &   0.514 &   2.569 &   0.590 \\
A0V  &   0.029 &   0.001 &   0.699 &   0.199 &   1.352 &   0.355 &   1.991 &   0.468 &   2.616 &   0.544 \\
A2V  &   0.046 &   0.023 &   0.716 &   0.220 &   1.370 &   0.374 &   2.009 &   0.486 &   2.634 &   0.560 \\
A3V  &   0.068 &   0.025 &   0.737 &   0.220 &   1.391 &   0.372 &   2.029 &   0.483 &   2.654 &   0.555 \\
A5V  &   0.107 &   0.035 &   0.776 &   0.228 &   1.428 &   0.379 &   2.065 &   0.488 &   2.689 &   0.558 \\
A7V  &   0.138 &   0.070 &   0.805 &   0.261 &   1.457 &   0.408 &   2.094 &   0.515 &   2.717 &   0.584 \\
F0V  &   0.212 &   0.114 &   0.876 &   0.299 &   1.525 &   0.441 &   2.159 &   0.542 &   2.780 &   0.606 \\
F2V  &   0.257 &   0.156 &   0.920 &   0.338 &   1.567 &   0.477 &   2.201 &   0.576 &   2.822 &   0.638 \\
F5V  &   0.263 &   0.185 &   0.926 &   0.364 &   1.574 &   0.502 &   2.208 &   0.600 &   2.830 &   0.661 \\
F8V  &   0.325 &   0.218 &   0.985 &   0.393 &   1.631 &   0.526 &   2.262 &   0.620 &   2.882 &   0.677 \\
G0V  &   0.368 &   0.204 &   1.028 &   0.378 &   1.673 &   0.510 &   2.303 &   0.602 &   2.921 &   0.658 \\
G2V  &   0.384 &   0.227 &   1.043 &   0.400 &   1.687 &   0.530 &   2.317 &   0.621 &   2.935 &   0.676 \\
G5V  &   0.394 &   0.239 &   1.052 &   0.411 &   1.695 &   0.540 &   2.323 &   0.630 &   2.941 &   0.684 \\
G8V  &   0.420 &   0.261 &   1.077 &   0.429 &   1.718 &   0.556 &   2.345 &   0.643 &   2.961 &   0.695 \\
K0V  &   0.445 &   0.278 &   1.098 &   0.443 &   1.736 &   0.567 &   2.360 &   0.651 &   2.973 &   0.700 \\
K2V  &   0.474 &   0.298 &   1.127 &   0.461 &   1.765 &   0.582 &   2.390 &   0.665 &   3.004 &   0.713 \\
K3V  &   0.548 &   0.303 &   1.199 &   0.463 &   1.836 &   0.581 &   2.460 &   0.660 &   3.073 &   0.705 \\
K4V  &   0.613 &   0.347 &   1.262 &   0.502 &   1.895 &   0.615 &   2.515 &   0.691 &   3.126 &   0.732 \\
K5V  &   0.650 &   0.390 &   1.301 &   0.544 &   1.937 &   0.658 &   2.561 &   0.735 &   3.175 &   0.778 \\
K7V  &   0.800 &   0.490 &   1.450 &   0.636 &   2.085 &   0.743 &   2.709 &   0.812 &   3.322 &   0.849 \\
M0V  &   0.903 &   0.499 &   1.553 &   0.641 &   2.188 &   0.743 &   2.811 &   0.807 &   3.424 &   0.839 \\
M1V  &   1.063 &   0.547 &   1.719 &   0.688 &   2.360 &   0.789 &   2.987 &   0.853 &   3.604 &   0.885 \\
M2V  &   1.144 &   0.624 &   1.795 &   0.756 &   2.431 &   0.849 &   3.055 &   0.907 &   3.669 &   0.934 \\
M3V  &   1.521 &   0.820 &   2.174 &   0.941 &   2.811 &   1.023 &   3.436 &   1.070 &   4.052 &   1.088 \\
M4V  &   1.829 &   0.889 &   2.470 &   0.995 &   3.096 &   1.061 &   3.710 &   1.095 &   4.315 &   1.099 \\
M6V  &   2.514 &   1.060 &   3.165 &   1.143 &   3.801 &   1.191 &   4.425 &   1.207 &   5.041 &   1.198 \\
 & & & & & & & & & & \\
O8III  &  -0.180 &   0.075 &   0.489 &   0.282 &   1.142 &   0.446 &   1.780 &   0.568 &   2.404 &   0.651 \\
B1-2III  &  -0.108 &   0.072 &   0.565 &   0.276 &   1.222 &   0.438 &   1.863 &   0.558 &   2.491 &   0.639 \\
B3III  &  -0.042 &   0.089 &   0.630 &   0.291 &   1.285 &   0.449 &   1.925 &   0.566 &   2.551 &   0.644 \\
B5III  &  -0.019 &   0.083 &   0.653 &   0.283 &   1.308 &   0.441 &   1.947 &   0.557 &   2.573 &   0.634 \\
B9III  &  -0.022 &   0.031 &   0.649 &   0.231 &   1.303 &   0.388 &   1.942 &   0.504 &   2.568 &   0.582 \\
A0III  &   0.013 &   0.022 &   0.682 &   0.220 &   1.335 &   0.375 &   1.973 &   0.489 &   2.598 &   0.565 \\
A3III  &   0.041 &   0.057 &   0.708 &   0.252 &   1.358 &   0.404 &   1.994 &   0.515 &   2.617 &   0.587 \\
A5III  &   0.107 &   0.089 &   0.772 &   0.280 &   1.420 &   0.429 &   2.053 &   0.537 &   2.674 &   0.606 \\
A7III  &   0.130 &   0.096 &   0.797 &   0.285 &   1.447 &   0.432 &   2.083 &   0.539 &   2.706 &   0.607 \\
F0III  &   0.177 &   0.103 &   0.844 &   0.290 &   1.497 &   0.435 &   2.134 &   0.539 &   2.759 &   0.606 \\
F2III  &   0.230 &   0.188 &   0.890 &   0.368 &   1.534 &   0.506 &   2.164 &   0.603 &   2.782 &   0.664 \\
F5III  &   0.269 &   0.177 &   0.930 &   0.356 &   1.575 &   0.492 &   2.206 &   0.588 &   2.825 &   0.648 \\
G0III  &   0.378 &   0.236 &   1.035 &   0.408 &   1.677 &   0.537 &   2.306 &   0.627 &   2.923 &   0.681 \\
G5III  &   0.449 &   0.279 &   1.102 &   0.444 &   1.739 &   0.567 &   2.364 &   0.651 &   2.977 &   0.700 \\
G8III  &   0.474 &   0.276 &   1.128 &   0.440 &   1.767 &   0.563 &   2.392 &   0.647 &   3.007 &   0.696 \\
K0III  &   0.505 &   0.284 &   1.157 &   0.445 &   1.793 &   0.564 &   2.417 &   0.646 &   3.029 &   0.692 \\
K1III  &   0.529 &   0.300 &   1.182 &   0.461 &   1.820 &   0.580 &   2.445 &   0.660 &   3.059 &   0.706 \\
K2III  &   0.574 &   0.325 &   1.225 &   0.481 &   1.860 &   0.596 &   2.483 &   0.673 &   3.095 &   0.716 \\
K3III  &   0.582 &   0.320 &   1.230 &   0.475 &   1.863 &   0.589 &   2.483 &   0.665 &   3.092 &   0.706 \\
K4III  &   0.670 &   0.351 &   1.320 &   0.502 &   1.955 &   0.611 &   2.578 &   0.683 &   3.191 &   0.721 \\
K5III  &   0.808 &   0.381 &   1.456 &   0.525 &   2.088 &   0.629 &   2.708 &   0.695 &   3.318 &   0.728 \\
M0III  &   0.935 &   0.473 &   1.581 &   0.611 &   2.212 &   0.708 &   2.830 &   0.768 &   3.438 &   0.796 \\
M1III  &   0.984 &   0.461 &   1.633 &   0.598 &   2.266 &   0.695 &   2.887 &   0.755 &   3.496 &   0.782 \\
M2III  &   1.040 &   0.512 &   1.679 &   0.641 &   2.304 &   0.730 &   2.916 &   0.782 &   3.519 &   0.803 \\
M3III  &   1.311 &   0.564 &   1.957 &   0.687 &   2.586 &   0.770 &   3.203 &   0.817 &   3.809 &   0.832 \\
M4III  &   1.652 &   0.644 &   2.296 &   0.753 &   2.924 &   0.822 &   3.539 &   0.857 &   4.144 &   0.860 \\
M5III  &   1.941 &   0.725 &   2.581 &   0.821 &   3.205 &   0.879 &   3.816 &   0.904 &   4.417 &   0.899 \\
\hline
\end{tabular}
\end{table*}

\begin{table*}
\caption{Synthetic tracks in the $(r' - H\alpha, r' - i')$ plane for supergiant
stars calculated for a range of reddenings.}
\label{cc_tracks_2}
\begin{tabular}{lrlllllllll}
\hline
Spectral & \multicolumn{4}{l}{Reddening:-} & & & & & & \\
Type     & \multicolumn{2}{l}{$E_{B-V} = 0.0$} & \multicolumn{2}{l}{$E_{B-V} = 1.0$} &
           \multicolumn{2}{l}{$E_{B-V} = 2.0$} & \multicolumn{2}{l}{$E_{B-V} = 3.0$} &
           \multicolumn{2}{l}{$E_{B-V} = 4.0$} \\
  & $(r'-i')$ & $(r'-H\alpha)$ & $(r'-i')$ & $(r'-H\alpha)$ & $(r'-i')$ & 
    $(r'-H\alpha)$ & $(r'-i')$ & $(r'-H\alpha)$ & $(r'-i')$ & $(r'-H\alpha)$ \\
\hline
B0I  &  -0.078 &   0.109 &   0.591 &   0.311 &   1.243 &   0.470 &   1.880 &   0.588 &   2.504 &   0.666 \\
B3I  &  -0.063 &   0.126 &   0.603 &   0.325 &   1.253 &   0.481 &   1.888 &   0.596 &   2.510 &   0.672 \\
B5I  &   0.029 &   0.147 &   0.693 &   0.341 &   1.340 &   0.491 &   1.973 &   0.601 &   2.594 &   0.672 \\
B8I  &   0.049 &   0.157 &   0.717 &   0.351 &   1.369 &   0.502 &   2.007 &   0.612 &   2.631 &   0.683 \\
A0I  &   0.060 &   0.135 &   0.726 &   0.327 &   1.377 &   0.477 &   2.012 &   0.586 &   2.636 &   0.657 \\
A2I  &   0.071 &   0.154 &   0.734 &   0.345 &   1.382 &   0.493 &   2.015 &   0.601 &   2.636 &   0.671 \\
F0I  &   0.191 &   0.131 &   0.856 &   0.318 &   1.505 &   0.462 &   2.140 &   0.565 &   2.762 &   0.631 \\
F5I  &   0.219 &   0.158 &   0.882 &   0.340 &   1.529 &   0.480 &   2.162 &   0.580 &   2.783 &   0.643 \\
F8I  &   0.248 &   0.183 &   0.908 &   0.363 &   1.553 &   0.500 &   2.184 &   0.597 &   2.803 &   0.657 \\
G0I  &   0.317 &   0.229 &   0.971 &   0.401 &   1.609 &   0.531 &   2.234 &   0.622 &   2.847 &   0.677 \\
G2I  &   0.357 &   0.241 &   1.010 &   0.412 &   1.648 &   0.540 &   2.272 &   0.629 &   2.885 &   0.682 \\
G5I  &   0.389 &   0.282 &   1.040 &   0.449 &   1.676 &   0.573 &   2.298 &   0.658 &   2.909 &   0.708 \\
G8I  &   0.471 &   0.312 &   1.109 &   0.468 &   1.731 &   0.584 &   2.340 &   0.661 &   2.940 &   0.704 \\
K2I  &   0.611 &   0.329 &   1.251 &   0.481 &   1.876 &   0.591 &   2.488 &   0.663 &   3.089 &   0.701 \\
K3I  &   0.680 &   0.384 &   1.316 &   0.528 &   1.935 &   0.632 &   2.543 &   0.699 &   3.140 &   0.733 \\
K4I  &   0.839 &   0.385 &   1.485 &   0.526 &   2.115 &   0.626 &   2.733 &   0.689 &   3.340 &   0.719 \\
M2I  &   1.316 &   0.521 &   1.956 &   0.642 &   2.580 &   0.723 &   3.193 &   0.769 &   3.795 &   0.784 \\
\hline
\end{tabular}
\end{table*}

    The effect of interstellar extinction on the sequences is illustrated 
in Fig.~\ref{tracks_red}.  With increasing reddening, the range of 
$(r' - H\alpha)$ colour spanned by each luminosity-class sequence
diminishes.  This shrinkage in dynamic range is due to effective wavelength 
of the $r'$ bandpass
lengthening, and moving closer to the $H\alpha$ bandpass as reddening becomes 
ever more extreme.  It may be seen in Fig.~\ref{tracks_red} that the 
supergiant sequence essentially reddens along itself, while the main and 
giant sequences shift across in a manner that sweeps out area in the 
colour-colour plane.  Also shown in this figure is the reddening locus for 
A0 dwarfs -- for non-degenerate stars and binaries this line amounts to an 
important boundary in that none should drop below it.  In practice, 
stars will be carried into the forbidden domain by a range of types of 
observational error.   Degenerate dwarfs and related objects with stronger
H$\alpha$ absorption than early A stars would also fall below this line.

\subsection{The IPHAS colours of emission line stars}

    To assess the impact on $(r'-H\alpha)$ and $(r'-i')$ colours of
increasingly strong H$\alpha$ emission, we have represented underlying 
stellar SEDs using simple power laws or, at later spectral types, 
blackbodies.  At the acceptable price of some approximation, such as ignoring 
stronger spectral features like the Paschen limit, this approach allows us to 
explore a broad range of SEDs flexibly and to quantify thresholds for the 
straightforward detection of H$\alpha$ emission.  We present 
results for 4 simple SEDs: three power laws of the form 
$F_\lambda \propto \lambda^{-\beta}$ with $\beta$ set equal to 4 
(Rayleigh-Jeans case, relevant to the hottest O stars), 3 (appropriate to 
$\sim$A0 stars) and 2.3 (the optically thick accretion disk case); a Planck 
function at a temperature of 5900~K that is a good match to the G2V SED in the 
P98 library.

\begin{figure*}
\begin{picture}(0,230)
\put(0,0)
{\includegraphics{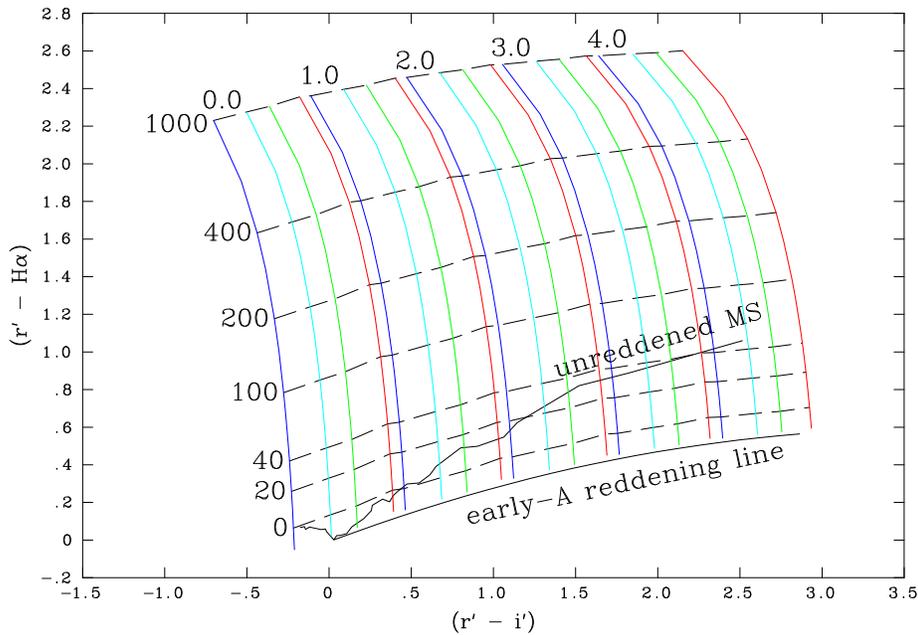}}
\end{picture}
\caption{The effect of adding in increasing H$\alpha$ emission to a
range of simplified stellar SEDs.  The black lines, representing an
unreddened main sequence and the early-A reddening trend, are the expected
bounds to the main stellar locus.  The dashed lines are lines of 
constant H$\alpha$ emission equivalent width, with the label at the
left specifying the equivalent width in Angstroms.  The vertical
coloured lines are, in effect, H$\alpha$ emission curves of growth for
particular choices of underlying SED and reddening: the darker blue lines are
for the Rayleigh Jeans case ($F_\lambda \propto \lambda^{-4}$) at different
reddenings; light blue and green lines are for power law indices of -3 and
-2.3 respectively, while the red lines are the results for a 5900~K
blackbody (a G2V star, roughly).  The lowest point on each curve corresponds 
to a narrow net H$\alpha$ absorption of EW, 10~\AA . The reddenings, as 
$E(B-V)$, are specified across the top and refer to each set of coloured 
lines.}    
\label{em_tracks}
\end{figure*}

\begin{table*}
\caption{Synthetic tracks in the $(r' - H\alpha, r' - i')$ plane, for 
increasing H$\alpha$ emission equivalent width.  Data are provided for
four underlying simplified stellar SEDs (3 power laws and a blackbody) and
for a range of reddenings.  The usual sign convention for H$\alpha$ EW
(left hand column) is reversed in that positive values refer to net 
emission.}
\label{em_colour}
\begin{tabular}{rrrrrrrrrrr}
\hline
H$\alpha$ & \multicolumn{4}{l}{Reddening:-} & & & & & & \\
EW (\AA ) & \multicolumn{2}{l}{$E_{B-V} = 0.0$} & \multicolumn{2}{l}{$E_{B-V} = 1.0$} &
           \multicolumn{2}{l}{$E_{B-V} = 2.0$} & \multicolumn{2}{l}{$E_{B-V} = 3.0$} &
           \multicolumn{2}{l}{$E_{B-V} = 4.0$} \\
 & $(r'-i')$ & $(r'-H\alpha)$ & $(r'-i')$ & $(r'-H\alpha)$ & $(r'-i')$ 
 & $(r'-H\alpha)$ & $(r'-i')$ & $(r'-H\alpha)$ & $(r'-i')$ & $(r'-H\alpha)$ \\
\hline
\multicolumn{6}{l}{1. $F_\lambda \propto \lambda^{-4}$ (Rayleigh Jeans):} & & & & & \\
       0 & -0.217 &  0.064 &  0.456 &  0.274 &  1.114 &  0.441 &  1.756 &  0.567 &  2.385 &  0.654 \\
      10 & -0.223 &  0.166 &  0.449 &  0.375 &  1.105 &  0.542 &  1.746 &  0.666 &  2.374 &  0.752 \\
      20 & -0.229 &  0.259 &  0.442 &  0.467 &  1.097 &  0.632 &  1.737 &  0.756 &  2.364 &  0.841 \\ 
      40 & -0.241 &  0.421 &  0.427 &  0.626 &  1.080 &  0.789 &  1.718 &  0.911 &  2.343 &  0.995 \\ 
      60 & -0.253 &  0.559 &  0.413 &  0.762 &  1.063 &  0.923 &  1.699 &  1.043 &  2.323 &  1.125 \\ 
      80 & -0.265 &  0.679 &  0.399 &  0.880 &  1.047 &  1.038 &  1.681 &  1.156 &  2.303 &  1.237 \\ 
     100 & -0.277 &  0.784 &  0.385 &  0.983 &  1.030 &  1.140 &  1.663 &  1.256 &  2.284 &  1.335 \\ 
     150 & -0.305 &  1.004 &  0.350 &  1.197 &  0.991 &  1.348 &  1.619 &  1.460 &  2.237 &  1.536 \\ 
     200 & -0.333 &  1.177 &  0.317 &  1.365 &  0.953 &  1.512 &  1.577 &  1.620 &  2.192 &  1.693 \\ 
     300 & -0.386 &  1.441 &  0.254 &  1.619 &  0.881 &  1.756 &  1.497 &  1.857 &  2.107 &  1.925 \\ 
     400 & -0.438 &  1.635 &  0.194 &  1.804 &  0.813 &  1.934 &  1.423 &  2.028 &  2.028 &  2.092 \\ 
    1000 & -0.702 &  2.232 & -0.109 &  2.363 &  0.477 &  2.460 &  1.061 &  2.528 &  1.647 &  2.573 \\
 1000000 & -7.094 &  3.241 & -6.632 &  3.241 & -6.143 &  3.242 & -5.627 &  3.242 & -5.086 &  3.242 \\ 
\hline
\multicolumn{6}{l}{2. $F_\lambda \propto \lambda^{-3}$ ($\sim$A0 SED)} & & & & & \\
       0 &  0.008 &  0.134 &  0.677 &  0.331 &  1.331 &  0.485 &  1.970 &  0.598 &  2.596 &  0.673 \\
      10 &  0.001 &  0.236 &  0.670 &  0.432 &  1.322 &  0.585 &  1.960 &  0.697 &  2.586 &  0.771 \\
      20 & -0.005 &  0.328 &  0.662 &  0.523 &  1.313 &  0.675 &  1.950 &  0.786 &  2.575 &  0.860 \\
      40 & -0.018 &  0.490 &  0.647 &  0.682 &  1.296 &  0.832 &  1.931 &  0.941 &  2.554 &  1.013 \\
      60 & -0.031 &  0.627 &  0.632 &  0.817 &  1.278 &  0.964 &  1.912 &  1.072 &  2.534 &  1.143 \\
      80 & -0.043 &  0.746 &  0.617 &  0.934 &  1.261 &  1.079 &  1.893 &  1.185 &  2.514 &  1.255 \\
     100 & -0.056 &  0.851 &  0.602 &  1.036 &  1.245 &  1.180 &  1.874 &  1.284 &  2.494 &  1.353 \\
     150 & -0.086 &  1.069 &  0.566 &  1.248 &  1.204 &  1.387 &  1.829 &  1.487 &  2.446 &  1.553 \\
     200 & -0.115 &  1.241 &  0.531 &  1.415 &  1.164 &  1.549 &  1.786 &  1.646 &  2.400 &  1.709 \\
     300 & -0.172 &  1.501 &  0.465 &  1.666 &  1.090 &  1.792 &  1.705 &  1.882 &  2.314 &  1.940 \\
     400 & -0.226 &  1.692 &  0.403 &  1.849 &  1.020 &  1.967 &  1.629 &  2.051 &  2.234 &  2.105 \\
    1000 & -0.503 &  2.277 &  0.088 &  2.396 &  0.674 &  2.484 &  1.260 &  2.544 &  1.849 &  2.583 \\
 1000000 & -6.940 &  3.241 & -6.468 &  3.242 & -5.969 &  3.242 & -5.444 &  3.242 & -4.893 &  3.242 \\
\hline
\multicolumn{6}{l}{3. $F_\lambda \propto \lambda^{-2.3}$ (optically thick disk accretion)} & & & & & \\ 
       0 &  0.165 &  0.181 &  0.832 &  0.368 &  1.483 &  0.513 &  2.120 &  0.617 &  2.744 &  0.684 \\
      10 &  0.158 &  0.283 &  0.824 &  0.469 &  1.474 &  0.613 &  2.110 &  0.716 &  2.733 &  0.783 \\
      20 &  0.151 &  0.375 &  0.816 &  0.560 &  1.465 &  0.703 &  2.100 &  0.805 &  2.723 &  0.871 \\
      40 &  0.138 &  0.535 &  0.800 &  0.718 &  1.447 &  0.859 &  2.080 &  0.960 &  2.702 &  1.024 \\ 
      60 &  0.125 &  0.672 &  0.784 &  0.853 &  1.429 &  0.991 &  2.060 &  1.090 &  2.681 &  1.154 \\
      80 &  0.112 &  0.791 &  0.769 &  0.969 &  1.411 &  1.106 &  2.041 &  1.203 &  2.661 &  1.265 \\
     100 &  0.099 &  0.895 &  0.754 &  1.071 &  1.394 &  1.206 &  2.022 &  1.302 &  2.641 &  1.363 \\
     150 &  0.067 &  1.112 &  0.717 &  1.282 &  1.352 &  1.412 &  1.977 &  1.504 &  2.593 &  1.563 \\
     200 &  0.037 &  1.283 &  0.681 &  1.448 &  1.312 &  1.574 &  1.933 &  1.662 &  2.546 &  1.719 \\
     300 & -0.022 &  1.541 &  0.613 &  1.697 &  1.236 &  1.814 &  1.850 &  1.897 &  2.459 &  1.949 \\
     400 & -0.078 &  1.730 &  0.548 &  1.878 &  1.164 &  1.988 &  1.773 &  2.065 &  2.379 &  2.114 \\
    1000 & -0.364 &  2.306 &  0.227 &  2.418 &  0.813 &  2.499 &  1.401 &  2.554 &  1.992 &  2.589 \\
 1000000 & -6.830 &  3.241 & -6.351 &  3.242 & -5.846 &  3.242 & -5.314 &  3.242 & -4.757 &  3.242 \\
\hline
\multicolumn{6}{l}{4. 5900 K blackbody ($\sim$G2 V SED)} & & & & & \\
       0 &  0.385 &  0.265 &  1.040 &  0.435 &  1.681 &  0.563 &  2.308 &  0.652 &  2.924 &  0.705 \\
      10 &  0.377 &  0.367 &  1.032 &  0.535 &  1.671 &  0.662 &  2.297 &  0.750 &  2.913 &  0.803 \\
      20 &  0.370 &  0.458 &  1.023 &  0.625 &  1.662 &  0.752 &  2.287 &  0.839 &  2.902 &  0.891 \\
      40 &  0.356 &  0.618 &  1.006 &  0.783 &  1.643 &  0.907 &  2.266 &  0.993 &  2.880 &  1.044 \\
      60 &  0.341 &  0.753 &  0.990 &  0.916 &  1.624 &  1.039 &  2.246 &  1.123 &  2.859 &  1.173 \\
      80 &  0.327 &  0.871 &  0.974 &  1.032 &  1.606 &  1.152 &  2.227 &  1.235 &  2.839 &  1.285 \\
     100 &  0.314 &  0.975 &  0.957 &  1.133 &  1.588 &  1.252 &  2.207 &  1.333 &  2.818 &  1.382 \\
     150 &  0.280 &  1.189 &  0.918 &  1.342 &  1.544 &  1.456 &  2.160 &  1.535 &  2.769 &  1.581 \\
     200 &  0.247 &  1.357 &  0.880 &  1.506 &  1.502 &  1.616 &  2.115 &  1.691 &  2.722 &  1.736 \\
     300 &  0.184 &  1.611 &  0.808 &  1.751 &  1.423 &  1.854 &  2.030 &  1.924 &  2.634 &  1.965 \\
     400 &  0.124 &  1.797 &  0.741 &  1.929 &  1.349 &  2.025 &  1.952 &  2.090 &  2.552 &  2.128 \\
    1000 & -0.177 &  2.358 &  0.406 &  2.456 &  0.988 &  2.526 &  1.571 &  2.572 &  2.161 &  2.599 \\ 
 1000000 & -6.695 &  3.241 & -6.210 &  3.242 & -5.698 &  3.242 & -5.161 &  3.242 & -4.598 &  3.242 \\
\hline
\end{tabular}
\end{table*}

   H$\alpha$ emission, where present, takes on a wide range of 
profiles in stellar spectra, and can be practically any width -- with FWHM
anywhere in the range from a few 10s of km s$^{-1}$ up to 1000s.  But, for now,
we treat the simple limiting case of H$\alpha$ emission that is 
well-contained within the width of the INT/WFC narrow-band H$\alpha$ filter.   
The particular realisation used is of a rectangular profile of 
breadth 25~\AA\ centred at 6570~\AA\ (note that the effective H$\alpha$ filter 
bandpass is blueshifted for objects observed off-axis, which is why the 
central wavelength of the WFC filter is accordingly specified as 6568~\AA ).  
This yields results negligibly different from using a Gaussian profile. 

   The synthesis of colours consists of the following steps: an underlying
stellar SED is chosen; a rectangular H$\alpha$ emission profile of the
desired equivalent width (EW) is superimposed; the resultant artificial 
spectrum is then reddened as required using the reddening law specified in 
section~\ref{normal}; finally, the reddened SED is multiplied by the product of
the survey filter profiles and WFC response and integrated to form colours as
in equation~\ref{eqn_colour}.  We have not attempted to apply this procedure
to very late type SEDs dominated by molecular bands -- in these stars, 
neither can the SED be easily parameterised, nor is an objective definition
of H$\alpha$ EW straightforward.  

   On the basis of this procedure we have synthesised colours for the same
set of reddenings ($E(B-V) = 0$ to $4$ in steps of $1$) for each of the
4 adopted stellar SEDs.  Our results are presented in Fig.~\ref{em_tracks}
and in Table~\ref{em_colour}.  We find that there is a practical degeneracy 
between reddening and underlying SED such that a very-nearly unique locus is 
traced at each adopted H$\alpha$ emission EW.  This means that, in principle, 
a given location in the colour-colour plane, above the main stellar locus, is 
associated with a particular H$\alpha$ EW.  

   A further property of the SED-specified tracks is that at EW up to 
$\sim$100~\AA , the trend with increasing EW is nearly vertical.  But as 
H$\alpha$ EW becomes very large, the tracks bend toward smaller $(r' - i')$ 
as the H$\alpha$ emission becomes a more significant contributor to the $r'$ 
flux.  Indeed beyond an EW of 1000~\AA\, as the switch from a `stellar' to a 
`nebular' spectrum with little discernable continuum takes place, the bending 
becomes very extreme.   In reality $(r' - i')$ in the nebular case will also
depend somewhat on the relative strength of line emission in the $i'$ band -- 
left out of consideration here.  The limiting value of $(r' - H\alpha)$ in the 
absence of any continuum is $\sim3.24$ for our synthetic system referred to 
the Hayes (1985) SED for Vega (see Table~\ref{em_colour}).  At the present 
time, without a properly defined zero point to the H$\alpha$ filter magnitudes,
this translates to an effective observed upper limit on $(r' - H\alpha)$ of 
around 3.1 -- any value appreciably above this signals a problem with the 
individual object's photometry.

   Finally an important feature to note in the trend in IPHAS colours with 
respect to both emission EW and reddening is that the threshold for the 
detection of H$\alpha$ emission is lowest for bluer and/or less reddened 
objects (see Fig.~\ref{em_tracks}).  This implies, for instance, that IPHAS 
will pick out faint, nearby accreting objects very well indeed down to just a 
few Angstroms EW.  Conversely, in the worst case of a densely populated main 
stellar locus spanning a wide range of reddenings, the EW threshold on the 
straightfoward detection of classical T Tau stars at $E(B-V) \sim 2$ 
($A_V \sim 6$) is around 30~\AA .  Not infrequently, however, at larger 
$(r' -i')$ ($\gtrsim 2$) the colour-colour plane below the unreddened main 
sequence may be sparsely populated -- the few objects located here could be
the result of a combination of anomalous reddening and line emission.  Indeed 
it is generally the case that objects, checked as having reliable photometry, 
lying outside the bounds of the densely populated main stellar locus for 
their field, have a relatively high probability of being interesting 
in one way or another.

\section{Simulated and observed IPHAS colour-colour diagrams compared -- 
fields in Aquila} 

\begin{figure}
\begin{picture}(0,340)
\put(0,0)
{\includegraphics{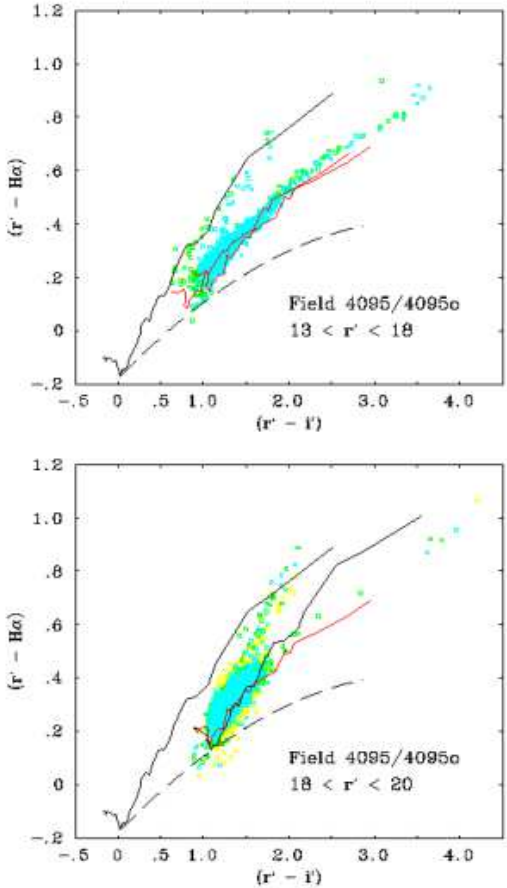}}
\end{picture}
\caption{The $(r' - H\alpha, r' - i')$ plane derived from observations
of IPHAS fiels 4095/4095o, compared with selected synthetic tracks.
In both panels, the uppermost solid line traces the unreddened MS track, while 
the dashed line is the early-A reddening line.  See text for specification of
other tracks.  In the upper panel, green points indicate $r' < 16$, while blue 
indicates $r' > 16$.  Below, yellow corresponds to $19.5 < r' < 20$, green to
$19 < r' < 19.5$ and blue to $18 < r' <19$.}
\label{4095_cc}
\end{figure}

    With the assistance of the synthetic tracks derived in the preceding
section it is possible to begin to make sense of the morphologies
appearing in IPHAS $(r' - H\alpha, r' - i')$ diagrams -- with a view to their
fuller exploitation. For this purpose we have selected some of the highest
quality IPHAS observations obtained from pointings in the Aquila Rift region,
allowing us to exploit its distinctive and easily interpreted colour-colour 
domain morphologies.

     We begin with field 4095 in Aquila, included in 
Fig.~\ref{aquila_cc} as the blue data points.  This field is roughly 
centred on $\ell = 32.5^{\rm o}, b = +4.8^{\rm o}$, sampling a region close to 
the survey's Galactic latitude upper limit.  The reddening data of Schlegel
et al (1998) indicates that $E_{B-V}$ typically does not exceed 1.6
in this direction.  This makes it the least obscured of the three fields and,
correspondingly, the field presenting the highest apparent density of 
stars (7097 out of the 13818 in Fig.~\ref{aquila_cc}).  In Fig.~\ref{4095_cc}
the data on extracted point sources are compared with selected synthetic 
tracks that have all been shifted downwards in $(r' - H\alpha)$ by 0.17 to
best match them to the data.  

     The upper panel in the Fig.~\ref{4095_cc} shows the brighter end 
of the magnitude range ($13 < r' < 18$) that includes a modest number of very 
nearly unreddened M dwarfs and a much larger number of mainly giant stars.  
Indeed the M giants form a particularly tight sequence at 
$(r' - i') \gtrsim 2.0$.  This suggests that most of the Galactic reddening 
along this sight line accumulates nearby because, if it were not, we would 
expect to see a more smeared giant distribution.  The simulated giant tracks 
for $E_{B-V} = 1.4$ and $1.6$ are compared with this very well-defined 
feature.  For $(r' - i') \lesssim 2$ many of the brighter objects will be 
giants at a plausible reddening; but at $(r' - i') > 2$, the synthesised 
tracks for M2-5~III stars fall too low by 
$\sim 0.05$ in $(r' - H\alpha )$.  A similar problem affects 
comparisons between synthesised and observed tracks for M dwarfs also 
(see below).  At $r' < 18$, only one object falls significantly below the 
early-A reddening line - it is likely to be a white dwarf or related object.

    The lower panel in Fig.~\ref{4095_cc} presents the faint end of the
$r'$ magnitude range, with the synthetic main sequence and giant tracks, 
reddened to $E(B-V) = 1.6$, superimposed.  The main locus of 
observed objects is now a little more steeply angled, indicating that these 
fainter stars include a much increased component of main sequence objects.  
However, at $r' \simeq 20$, stars later in spectral type than mid-K are only 
detectable at $E(B-V) \lesssim 0.8$, as evidenced by the scatter of points 
extending the main stellar locus up to $(r' - H\alpha) \sim 0.9$.  The 
small proportion of the plotted objects falling below the early-A reddening 
line can be presumed consistent with observational error.  The 0.17 offset of 
the synthesised tracks was determined by optimising the positioning of both 
this notional line and the unreddened main sequence with respect to the data 
for $r' \leq 18$.  These particular IPHAS observations have captured objects 
out to the limits of the Galactic disc population, such as reddened mid-M 
giants at $\sim 10$~kpc, located around 800~pc above the mid-plane at about 
the location of the far Sagittarius-Carina arm.

\begin{figure}
\begin{picture}(0,170)
\put(0,0)
{\includegraphics{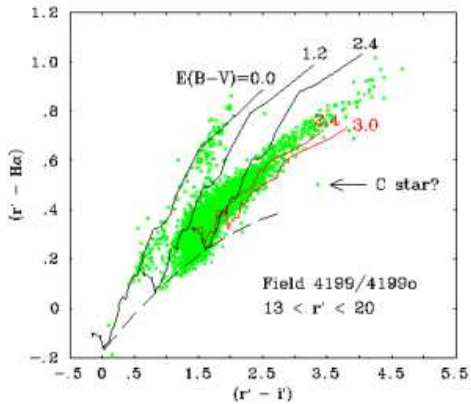}}
\end{picture}
\caption{The $(r' - H\alpha, r' - i')$ plane derived from observations
of IPHAS fields 4199/4199o, compared with selected synthetic tracks.  The
solid black lines are main sequence tracks for $E(B-V) = 0$ and 2.4,
while the dashed line is the early-A reddening line.  The two red tracks are
giant tracks, for $E(B-V) = 2.4$ and 3.0. The position of the probable C star
discussed in the text is picked out by the arrow.}
\label{4199_cc}
\end{figure}

    The more highly reddened Aquila field 4199, centred at $\ell = 
34.3^{\rm o}$, $b = +1.8^{\rm o}$, provides some degree of contrast with 
4095 and is illustrated in Fig.~\ref{4199_cc}.  The reddening here is 
more variable with position, as well as more extreme.  This shows itself
directly in the broader red giant locus.  Superimposed on the colour-colour
plot are synthesised giant-star tracks for $E_{B-V} = 2.4$, and $3.0$.  The 
latter value was selected because the maximum Galactic extinction for this 
field, derived from the Schlegel et al 1998 mapping data, 
is $E_{B-V} \simeq 3.0$.  There is a rough consistency here with the findings 
from field 4095, in that the M giant track synthesised for this extinction 
falls a bit below the observed thinning of putative M giants (as it did for
field 4095).  Down to $r = 20$, the colour-colour data suggest the presence of 
main sequence stars of K and earlier type at reddenings in the range 
$1.2 \lesssim E(B-V) \lesssim 2.4$: the main sequence tracks for these limits 
are drawn to illustrate this (Fig.~\ref{4199_cc}).  This field again bears 
the imprint of the Aquila Rift in the relative deficit of stars between the 
lightly populated zero-extinction main sequence and the dense locus of stars 
at $E_{B-V} \gtrsim 1.2$.  

   In field 4199 it is more apparent that the observed unreddened M dwarfs 
tend to maintain the locus gradient defined at earlier spectral types, rather 
than begin to turn over as the synthesised track indicates they should.  This 
is most likely another symptom of the problem behind the M-giant discrepancy.  
At the present time, the available conversion between Landolt $(R - I)$ 
colours, appropriate to the standard star fields, and Sloan $(r' - i')$ are 
not properly defined for M dwarf colours (see Smith et al 2002).  Similarly, 
the existing conversion used in the CASU pipeline is not validated for 
$(r' - i') > 1.5$.  Clearly this will need to be corrected in the future.  In 
the mean time, the comparison between observations of M stars and synthesised 
data will be increasingly qualitative as $(r' - i')$ increases beyond the 
validation limit.

   In neither field 4095 nor field 4199 do supergiants stand out in any obvious
morphological way.  This is likely to be both a consequence of their 
relative rarity and of the way in which their locus shifts almost along 
itself with increasing reddening.  In principle, extremely red, isolated 
objects located below the red giant locus could be picked out as candidate
reddened supergiants -- or as potential examples of other interesting
object types.  Indeed, we find that IPHAS J184644.25+015324.6, the one 
isolated object in this part of the 4199 colour-colour plane at $(r' - i', 
r' - H\alpha) = (3.36\pm0.02, 0.499\pm0.03)$, cannot be a reddened 
supergiant.  One reason is that the reddening ($E(B-V) \gtrsim 4$) 
required to explain its position in these terms is excessive relative to the 
maximum expected for the field.  Another is that the 2MASS point source 
within 0.2 arcsec of this object's position exhibits very bright 
$JHK$ magnitudes with unusual colours ($K = 9.55 \pm 0.02$, $(J - H) = 
1.68 \pm 0.03$, $(H - K) = 0.76 \pm 0.03$).  At $r' = 18.75 \pm 0.01$, 
there are no grounds for doubting the reliability of the IPHAS photometry 
and the reality of the source.  The absence of any significant proper motion 
rules this object out as a nearby brown dwarf.  This object is largely 
absent from pre-existing photographic surveys, except that there is a 
detection of it in the UKST Infrared (IVN) Survey reported on the SuperCOSMOS 
Sky Survey website (http://www-wfau.roe.ac.uk/sss/): it is reported there at 
an $I$ magnitude of 18.383, around 3 magnitudes fainter than the IPHAS $i'$ 
magnitude. The IVN plate was obtained in 1981. This variability, the 
anomalously low $(r' - H\alpha)$ colour, together with its NIR colours, point 
towards a carbon star at a reddening corresponding to $E(B-V) \sim 1.4$ 
(see e.g. Bessell \& Brett 1988).  A faint-end absolute $K$ magnitude for such
a star, if on the AGB, would be -6.5 (Claussen et al 1987).  This places it 
at $\sim$15~kpc.

\section{A comparison between flux-calibrated spectra and IPHAS photometry -- a
field in Taurus}

\begin{table*}
\caption{Observed and derived properties of the follow-up sample of stars
in IPHAS fields 2540/2540o in Taurus.} 
\label{2540_stars}
\begin{tabular}{lclrrlrrrl}
\hline
   & IPHAS name/position & \multicolumn{3}{c}{IPHAS photometry} 
& spectral & $E_{B-V}$ & \multicolumn{2}{c}{synthetic colours} & comment 
\\ 
   & J[RA(2000)$+$Dec(2000)] & $r'$ & $r'-i'$ & $r'-H\alpha $ & type &   
   & $r'-i'$ &  $r' - H\alpha $ & \\
\hline  
A & J053430.11$+$251400.9 & 13.19 & 0.33 & -0.07 & A2V & 0.40 & 0.32 & 
 0.09 & \\
B & J053432.14$+$252231.0 & 15.82 & 0.54 &  0.12 & A0V & 0.65 & 0.47 & 
 0.26 & H$\alpha$ emission \\
C & J053425.99$+$250843.1 & 13.01 & 0.61 &  0.16 & G0III & 0.35 & 0.61 & 
 0.30 & \\
D & J053458.93$+$252316.7 & 17.22 & 1.91 &  0.78 & M4V & 0.05 & 1.86 & 
 0.92 & \\
E & J053311.04$+$251444.5 & 15.36 & 1.15 &  0.30 & G0III & 1.10 & 1.11 & 
 0.42 & \\
F & J053305.64$+$251837.6 & 17.06 & 0.86 &  0.13 & A5 & $\sim$1.1 & 0.84 & 
 0.25 & noisy ISIS data\\
\hline
\end{tabular}
\end{table*}

\begin{figure*}
\begin{picture}(0,230)
{\includegraphics{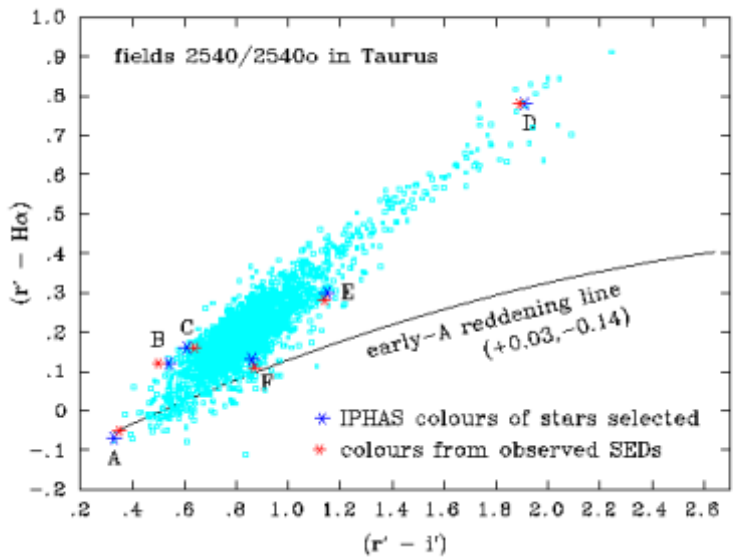}}
\end{picture}
\caption{The $(r'-H\alpha, r'-i')$ diagram for IPHAS fields 2540 and 2540o,
sampling an area roughly equivalent to 30$\times$30 sq.arcmin around 
RA 05 33 40, Dec +25 20 00 (J2000) in Taurus.  The magnitude range shown is
$13 \leq r' \leq 20$.  The dark blue asterisks mark the colours of the 6 
objects for which WHT/ISIS service spectra were obtained (stars A--F, see
also Fig.~\ref{2540_spec} and Table~\ref{2540_stars}).  The red asterisks 
mark the `predicted' colours derived from the flux-calibrated 
spectra, after applying the same shift to all 6 objects in order to minimise 
the mean difference with respect to the original IPHAS colours.  Like the 
computed colours for stars A--F, the synthesised early-A reddening line has 
been shifted by $+0.03$ and $-0.14$ in $(r' - i')$ and $(r' - H\alpha)$, 
respectively.}
\label{2540_cc}
\end{figure*}

Up to this point the interpretation of the $(r' - H\alpha, r' - i')$ plane has
been based on synthetic photometry derived from P98 library spectra.  Early 
in February 2004, we obtained WHT/ISIS service spectra of 6 stars selected
from IPHAS data on fields 2540 and 2540o in Taurus.  These IPHAS images, 
obtained on 5th November 2003, were among the first to be pipeline-processed 
and were picked for closer investigation as examples of apparently good 
quality data obtained in good seeing and photometric conditions.  The aim of 
the follow-up service spectra was to obtain relative spectrophotometry in 
order to ascertain optical SEDs, spectral types and reddenings for the sample 
stars as a retrospective check on the IPHAS colours, and the typical errors in 
them.  This exercise gives an impression both of the current state of the 
photometric calibration of the data and of the quality of the synthetic colour 
comparisons derived from P98.

Field 2540 is centred on RA 05 33 49 Dec $+$25 15 00 (2000) in 
Taurus, only a few degrees from the Galactic anticentre direction.  The 
characteristics of this sky position are very different from those in Aquila: 
here, the maximum Galactic reddening is modest and more smoothly varying, 
ranging from $E(B-V) \simeq 0.8$ in the NE of the $\sim 30\times30$ arcmin$^2$
extracted region up to $\sim 1.1$ in its SW.  The other obvious difference, 
which stands out in the $(r' - H\alpha, r' - i')$ plot for 2540/2540o in 
Fig.~\ref{2540_cc}, is the absence of any red giants within the magnitude 
range 
shown ($13 < r' < 20$).  This absence is not just a consequence of the imposed 
magnitude limits since a K/M giant at 10~kpc viewed through $\sim$3 visual 
magnitudes of extinction should be detected at $r' \sim 17$: it must be a real 
absence.  In the example of solar and lower metallicity isochrones presented 
by Bertelli et al (1994), a red giant branch is only well-developed from 
around 100 million years of age onwards -- suggesting that the stellar 
populations sampled in this part of the outer Galaxy are younger than this.

\begin{figure*}
\begin{picture}(0,440)
\put(0,0)
{\includegraphics{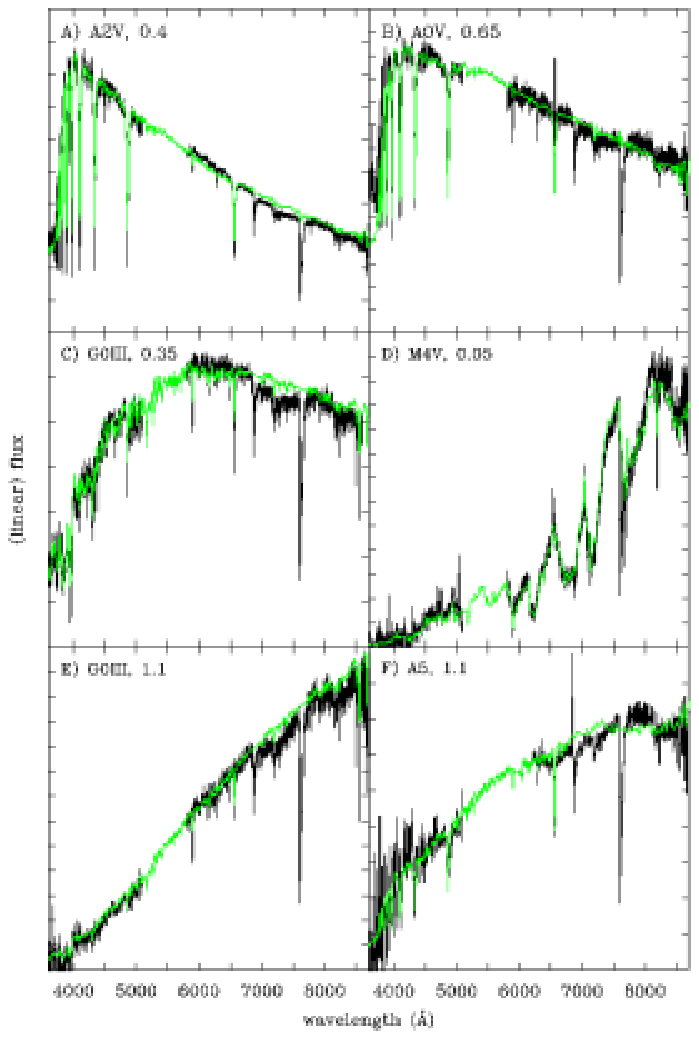}}
\end{picture}
\caption{Fits of P98 library spectra (green) to flux-calibrated 
WHT/ISIS spectra (black) of selected stars in IPHAS field 2540/2540o in 
Taurus.  The label in each box (letters A to F) refer to the positions marked 
in the colour-colour diagram for this field (Fig.~\ref{2540_cc}). Next to each 
label is the spectral type and colour excess ($E(B-V)$) estimate for the star.
In each plot, the flux scale is linear, with zero as the minimum plotted 
value.  The ISIS spectra are uncorrected for telluric absorption.}
\label{2540_spec}
\end{figure*}

8 stars were initially selected for ISIS spectroscopy from the colour-colour 
diagram for fields 2540/2540o on the basis that they were not too faint 
($r' < 18$) and lay on the outer boundary of the main stellar locus (dark blue 
asterisks in Fig.~\ref{2540_cc}).  These criteria biased the selection in 
favour of evolved spectral types.  In the event, 6 of the 8 stars (which we 
refer to as stars A--F) were 
observed during a service night of mediocre weather.  Both the blue and red 
arms of ISIS were used, with the R600B and R316R gratings installed, 
delivering spectra spanning 3500--5000~\AA\ and 6000--8700~\AA\ for each star. 
To further the aim of relative spectrophotometry, the slit width was set 
fairly wide at 1.8 arcsec, while the slit orientation tracked the parallactic 
angle.  The resolution of the spectra is $\sim3.6$~\AA .  An 
observation of the white dwarf G191$-$B2B was also obtained to serve as a 
spectrophotometric flux standard.  

The data were extracted from the CCD frames and then wavelength- and 
flux-calibrated using routines from the software package, FIGARO.  The 
extracted 1-D spectra were then imported to the software tool, DIPSO, in order 
to determine approximate spectral types and reddenings by comparing them with 
P98 library spectra.  In every case, the spectral type determination rested on 
matching absorption line characteristics.  This matching was performed using 
the blue spectra for all but the M4V star (star D) -- for this object the red 
spectrum was more appropriate.  For each star, the P98 library spectrum 
of the appropriate spectral type was progressively reddened, using the mean 
Galactic extinction law (Howarth 1983), to identify the best fitting colour 
excess.  The observed spectra and the best fits to them derived in this manner 
are shown in Fig.~\ref{2540_spec}.  The positions, magnitudes and further 
data on the 6 stars appear in Table~\ref{2540_stars}.  For star F, the 
data are not of sufficient quality to provide a reasonable fix on luminosity 
class: but we suspect that its H$\alpha$ profile indicates a lower gravity 
than a main sequence A5 star.  In order to: smooth errors due to irregularities
in the spectrophotometric flux calibration; avoid the need to correct for 
telluric absorption; provide a good extrapolation of the observed spectral 
energy distributions (SEDs) to cover the full spectral range, the final step 
of deriving photometric colours used the closest matching, 
appropriately-reddened P98 library spectra in the $r'$ and $i'$ 
bands, rather than the calibrated observations.  Only the H$\alpha$ 
fluxes were computed by multiplying the H$\alpha$ filter profile directly with 
the calibrated ISIS spectra.

Both the original IPHAS colours and the colours derived from the fits to the
spectrophotometry are listed in Table~\ref{2540_stars}.  As there is not
yet a uniform and fully-verified zero-point calibration for all IPHAS frames,
we have to shift the SED-based colours on to the IPHAS
colours.  The shifts that minimise the mean differences in each of 
$(r' - H\alpha)$ and $(r' - i')$ are $-0.14$ and $+0.03$, with final rms 
deviations between the 6 pairs of observed and predicted colours of 0.015 and 
0.026, respectively.  The size and sense of shift in $(r' - H\alpha)$ is as 
expected (see section~\ref{simul}).  That the shift in $(r' - i')$ is small, 
but apparently finite, indicates that the night the IPHAS imaging was obtained 
was not perfectly photometric.  On the basis of the errors estimated for the 
IPHAS photometry (see Fig.~\ref{errors}), we would expect the rms 
deviations between the catalogued and observed colours to be $\sim$0.01.  The 
somewhat larger values of 0.015 and 0.026 obtained here turn out to be 
determined mainly by errors in the relative spectrophotometry and its 
analysis: for example, the uncertainty in the $E(B-V)$ estimates is typically 
0.05 and translates into a $(r' - i')$ error of $\sim$0.03.  In 
$(r' - H\alpha)$, the discrepancies are smaller and mainly arise in the $r'$ 
band integration.  The circle from photometry to spectroscopy, back to
photometry, closes satisfactorily.

\begin{figure}
\begin{picture}(0,360)
\put(0,0)
{\includegraphics{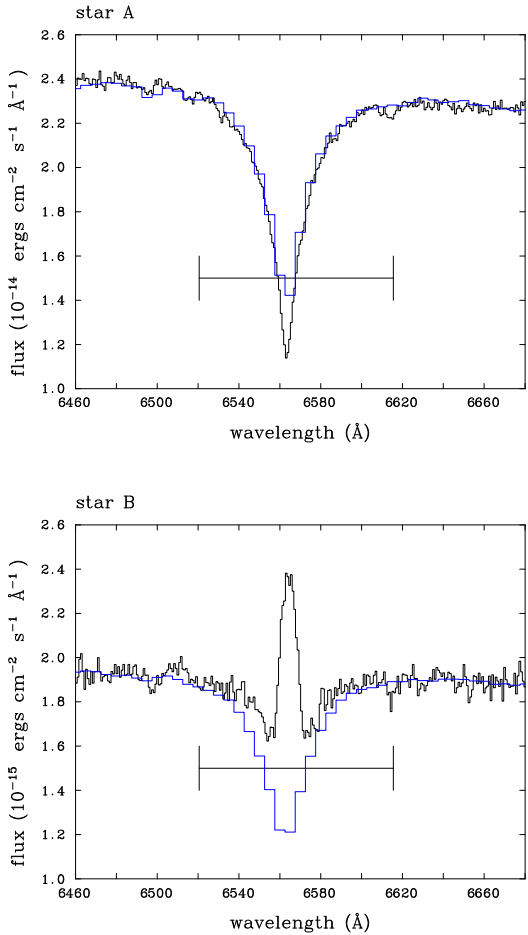}}
\end{picture}
\caption{Excerpts of the spectra obtained for stars A and B, focusing on the
H$\alpha$ line profiles.  The WHT/ISIS observations are plotted in black, 
while the
P98 library spectra of the appropriate spectral types (respectively A2V and
A0V) are shown superimposed in blue.  The excess emission equivalent width
in star B is 12~\AA .  In both panels the positioning and FWHM of the IPHAS 
H$\alpha$ filter is indicated by the horizontal bar.}
\label{2540_ex}
\end{figure}

\begin{figure}
\begin{picture}(0,190)
\put(0,0)
{\includegraphics{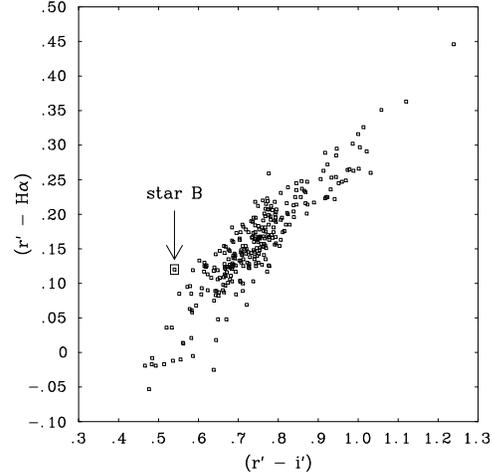}}
\end{picture}
\caption{The $(r' - H\alpha$, $r' - i')$ plane in the locality of star B.
The plotted objects are selected with $13 < r' < 19$ and are located in a
10$\times$10 arcmin$^2$ sky area centred on RA 05 33 40 Dec $+$25 20 00 (2000).
In the absence of its weak H$\alpha$ emission, star B would have fallen 
0.13 lower in $(r' - H\alpha)$, at a position typical of non-emission A0V 
stars.}
\label{2540_starB}
\end{figure}

At the end of this process, it can be seen in Fig.~\ref{2540_cc} that star A 
(A2V) falls a little below the early-A reddening line, rather than just above, 
while the reddening line itself lies $\sim 0.02$ magnitudes above the bottom 
edge of the main stellar locus.  A part of the reason for this may be 
illustrated in Fig.~\ref{2540_ex} where the P98 A2V spectrum is superimposed 
on the ISIS observation of star A: it is possible that the limited resolution 
of the P98 library spectra ($R \simeq 500$ or $\Delta \lambda \simeq 13$~\AA\ 
at H$\alpha$) leads to the H$\alpha$ in-band fluxes of early A-type stars 
being overestimated, very slightly.  Another factor will be linked to the 
question of the mean H$\alpha$ absorption EW and its variance for early-A 
stars as a function of sub-type and metallicity.  The H$\alpha$ absorption EW 
for star A is $14.5 \pm 0.4$~\AA , while that for the P98 A2V star is 
$10.8 \pm 0.5$.  These numbers for A2V may be compared with the P98 A0V and 
Hayes (1985) Vega H$\alpha$ EWs, that are both close to 13.0~\AA . Finally we 
note the contrast between the statement by Jaschek \& Jaschek (1987) that the 
Balmer lines are strongest at A2 (see their Table 10.1) and the maximum in P98 
for near-solar metallicity dwarfs at A0.  The underlying practical difficulty 
here is the measurement and calibration challenge of the very broad 
H$\alpha$ absorption wings in A-type spectra.  As the IPHAS survey completes 
and a uniform photometric calibration is constructed, it will then be 
appropriate to sort the issue out and improve the absolute registration 
of this lower boundary.  For the present, the A0V spectrum in the P98 library 
defines the shape of the early-A reddening line well enough to allow it to be 
used in a relative manner. 

The H$\alpha$ profile for star B is also presented in Fig.~\ref{2540_ex}.
This object shows a distinct central reversal in H$\alpha$ which allows it
to be described as a weak emission line object.  The excess (emission) 
equivalent width with respect to the library A0V spectrum is 12~\AA .  Indeed 
it was included in the list of ISIS service targets because of the suspicion 
that this might be the case: when it is plotted on the colour-colour plane 
with only neighbouring stars, within a 10$\times$10 arcmin$^2$ box, it sits 
clearly separated in $(r' - H\alpha)$ just above the local main stellar locus 
(see Fig.~\ref{2540_starB}).  This is an example of the greater coherence of 
mean conditions (reddening, nature of population) within a smaller sky area 
leading to greater success in identifying an `unusual' object.

\section{Spectroscopic trawling for emission-line and other rare objects -- 
fields in Cepheus}

\begin{figure}
\begin{picture}(0,370)
\put(0,0)
{\includegraphics{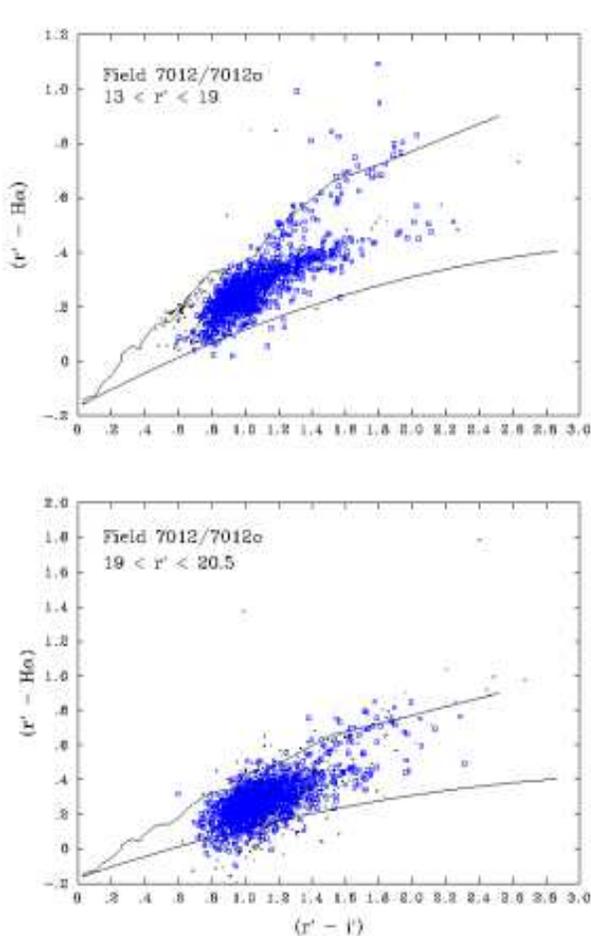}}
\end{picture}
\caption{Point-source colour-colour data derived from IPHAS fields 7012/7012o
in Cepheus.  This plot illustrates the quality of data around
($\ell = 105.6^{\rm o}, b = +4.0^{\rm o}$) from which we selected targets 
for multi-fibre spectroscopy with MMT/HectoSpec.  The magnitude range from
which HectoSpec targets were selected was $(17 < r' < 20)$ (blue data
points in both panels -- data outside this magnitude range are plotted as
smaller black points).  The black lines in both panels are the simulated 
unreddened main sequence track and early-A reddening line, shifted downward 
in $(r' - H\alpha)$ by 0.16 magnitudes.  Here they serve as fiducial lines to 
show how the errors grow at fainter magnitudes.}
\label{7012_cc}
\end{figure}

We now present some results from early spectroscopic follow-up of 
IPHAS in order to give a concrete example of the yields of different object 
types from IPHAS data.  In this case, the choice of sky area has been 
dictated mainly by observational convenience, rather than by data
quality considerations.  The results presented here rest on more typical
IPHAS photometry.  

\subsection{The MMT/HectoSpec observations}

In June 2004 we obtained spectra with the Mount Hopkins 6.5-metre MMT in 
F/5 configuration using the recently commissioned HectoSpec facility, 
a multi-object spectrograph fitted with 300 fibres that can be deployed across 
a field, 1 degree in diameter (Fabricant et al 2004).  The fibre positioner is 
mounted at Cassegrain.  The 270 groove/mm grating used delivers broad 
wavelength coverage (4488 -- 8664 \AA ) at 6.2~\AA\ resolution. Over two
nights, six different fields were observed using two fibre configurations 
per pointing.  The target stars selected for this programme fell 
mainly in the magnitude range $17 \leq r' \leq 20$.   The total
on-source exposure times were 1200 secs.  Spectra were extracted by the 
instrument pipeline that includes CCD bias and gain corrections, flat-fielding 
using domeflats as well as a sensitivity correction for the individual fibres 
using twilight flats. Individual fibre spectra were then extracted and 
wavelength calibrated using FeNeAr-lamp exposures. Finally, a mean sky 
spectrum derived from the sky fibres was subtracted.  Due to spatial 
variations in the sky background across the field of view, sky subtraction 
using sky fibres is not always perfect and care must be taken with spectra 
displaying weak, unresolved H$\alpha$ emission components. Sky subtraction 
will be improved in the future using offset sky exposures that sample the sky 
for each fibre close to the target position, in conjunction with an optimal 
scaling correction using the strongest sky lines.

\subsection{Selecting targets for the Cepheus field}

We report the results from one of two pointings in the constellation of 
Cepheus.  Centred on RA 22 17 00, Dec $+$61 33 37 (2000) 
($\ell = 105.6^{\rm o}$, $b = 4.0^{\rm o}$), this position was picked because 
it contains a strip included in the 
Spitzer Galactic First Look Survey (http://ssc.spitzer.caltech.edu/fls/galac/).
Despite its relatively high Galactic latitude, this area of sky presents 
significant and locally-variable interstellar extinction (ranging from 
$A_V \sim 4$ up to $\sim$7 magnitudes).  This shows up in the IPHAS colour 
data for the region as somewhat broadened main stellar loci.  An example 
is shown as Fig.~\ref{7012_cc}, where colour data extracted from the 
IPHAS field pair, 7012/7012o, are plotted.  At magnitudes brighter than 
$r' \sim 19$ (top panel), the colour uncertainties are less than $\sim$0.05 
and are less significant than environmental factors in the smearing of the 
main stellar locus -- this reverses at fainter magnitudes (lower panel) where 
the typical errors are $\sim0.1$.  

\begin{figure}
\begin{picture}(0,200)
\put(0,0)
{\includegraphics{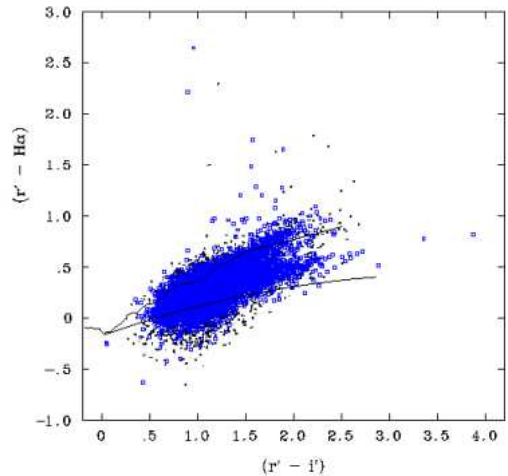}}
\end{picture}
\caption{The point-source colour-colour diagram used for target selection in 
the 1$^{\rm o}$-diameter field in Cepheus observed with MMT/HectoSpec.  Data 
for objects in the magnitude ranges $17 < r' < 20$ and $20 < r' < 20.5$ are 
shown in blue and black respectively.  Because of IPHAS field merging and the 
accompanying ad hoc photometric corrections, the main stellar locus is more 
diffuse here than in individual field extractions (cf Fig.~\ref{7012_cc}).  
The simulated unreddened main sequence track and the early-A reddening line 
(drawn as solid lines) are accordingly less useful in decoding the data.}  
\label{cepheus_cc}
\end{figure}

The HectoSpec 1$^{\rm o}$-diameter field spans then IPHAS field positions, not 
including offsets.  The largest contributions, however, are from IPHAS fields 
6985, 6993, 7012 and 7019 (see Table~\ref{field_list}).  To speed up the 
compilation of the target lists, we chose to merge the data from all the 
relevant pointings first, before proceeding to target selection.  In merging 
the data, corrections for photometric shifts between different WFC exposures 
had to be applied.  These were calculated using the mean magnitude 
offsets for sources located in field overlaps.  Inevitably, this merging 
blurred the main stellar locus in the colour plane some more -- compare the 
lower panel in Fig.~\ref{7012_cc} with the plot for the full HectoSpec field 
in Fig.~\ref{cepheus_cc}.  For the future, we are re-ordering the algorithm, 
in order to give more emphasis to selection at the individual field level (cf 
the discussion of 'star B' at the end of Section 5).  Indeed a further tactic 
that can be applied in order to minimise the spread of the main stellar locus 
is to select from within a number of narrowly-set $r'$ magnitude ranges.   On 
this occasion, after merging the catalogues for the relevant individual fields 
together, the main target selection was performed within the full magnitude 
range to be observed, $17 \leq r' \leq 20$.  Finally a few promising emission 
line star candidates down to $r' = 20.5$ were added by hand. 

The goal of this first round of HectoSpec observations was to explore 
the complete IPHAS colour-colour plane while trying to give high priority to 
objects that are outliers or near the edge of the general distribution of 
objects. This naturally includes all emission line star candidates, which lie
above the main stellar locus.  

To achieve this sampling, the following selection algorithm was applied.
The colour-colour plane was split into bins of 0.1 magnitudes in width and
height. Targets were then selected based on the number of objects in each bin:
from 1 to 3 objects in a bin, all were selected and given the highest fibre
allocation priority; between 4 to 9 objects in a bin, a random selection of 
75\% of the bin members was chosen with a slightly lower fibre allocation
priority; between 10 and 50 objects, a random fraction falling linearly
from 50\% to 10\% was selected and given a low fibre selection priority based 
on the number of objects in the bin; $>$50 objects in a box, 10\% of the
objects were selected randomly and given a low fibre selection priority based 
on the number of objects in the bin. The maximum number of objects that could 
be selected in a box was capped at 10.  Finally, if a bin had 4 neighbouring 
bins in the cardinal directions with 10 or more objects, then only one object 
was selected for spectroscopy.  This rule was introduced to further reduce the 
number of objects picked for spectroscopic follow up that lie in the densest 
-- and probably most uninteresting -- parts of the stellar distribution.

\begin{figure*}
\begin{picture}(0,370)
\put(0,0)
{\includegraphics{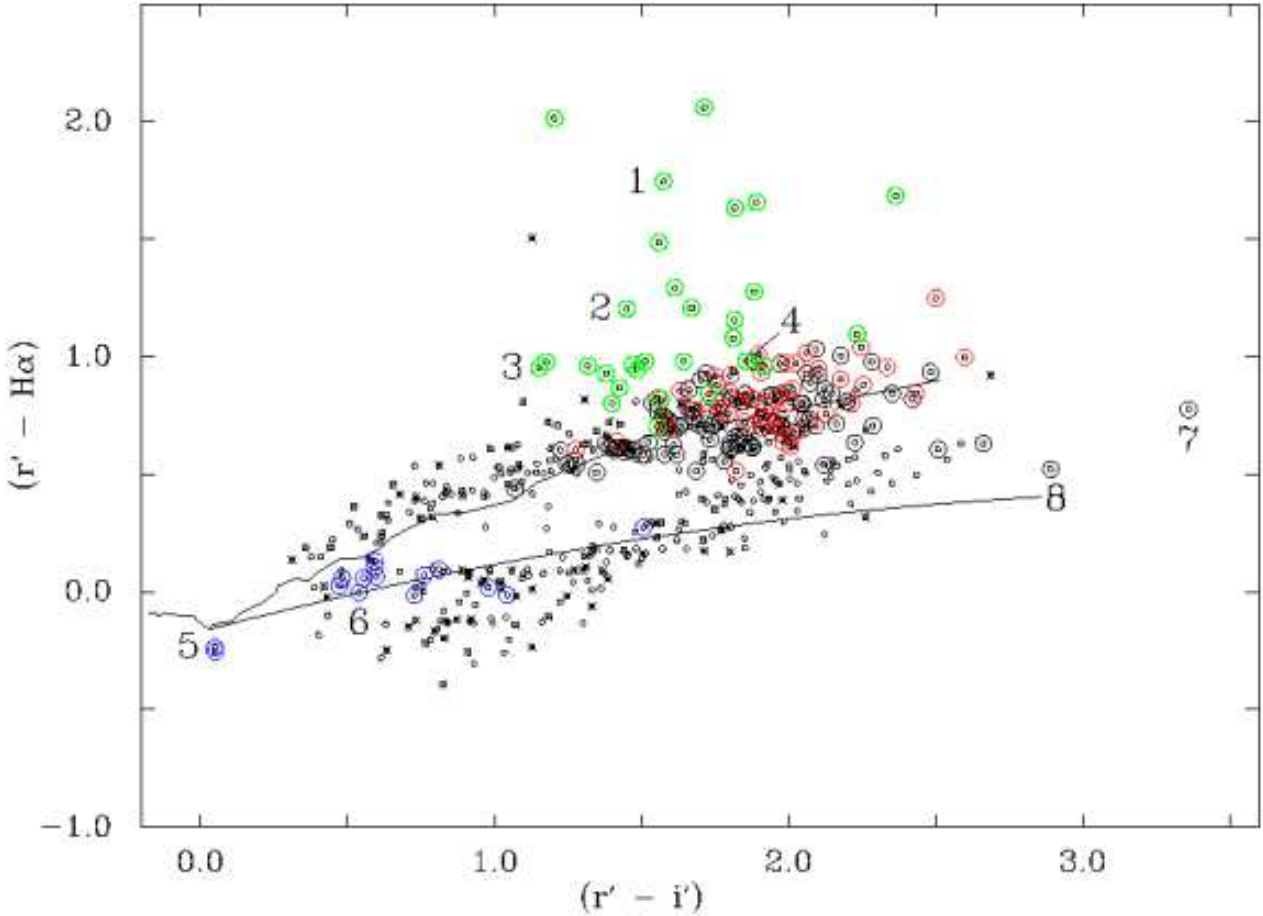}}
\end{picture}
\caption{The colour-colour diagram of the Cepheus targets observed with 
MMT/HectoSpec.  Symbols used: each small black circle locates the IPHAS 
colours of an object for which we have a spectrum; larger green circles pick 
out confirmed emission line stars; red circles pick out weaker-emission, 
possible dMe stars; larger black circles indicate non-emission line late-type 
stars with strong molecular bands in their spectra; blue circles pick out 
stars with prominent H$\alpha$ and H$\beta$ absorption features; black crosses 
mark those objects for which the spectrum is too noisy to classify.  Small 
black circles left unadorned will mainly represent late-A to early K stars.  
The synthetic tracks, drawn as solid lines are as in Figs.~\ref{7012_cc} and 
\ref{cepheus_cc}.  Spectra of the objects 1--8 are shown in 
Figs.~\ref{rogues_1} amd \ref{rogues_2} and parameters on them are given in
Table~\ref{hec_data}.
}
\label{hec_cep}
\end{figure*}

If the selection algorithm gave less than the requested number of objects
($\sim 700$ for 2 HectoSpec configurations) then 1 object per bin was 
added, starting with the least populated bins until the required number of 
objects was reached. If more than the requested number of objects was chosen, 
then 1 object was removed from each bin with more than 1 object selected 
-- starting with the most populated bin until the requested number of objects 
was reached.  By this means we were able to assign fibres to between 450 and 
540 targets, depending on the field observed.  Each set was then split into 
two configurations.

The outcome of this process and the HectoSpec observation of the Cepheus field 
was a collection of 496 stellar spectra. 

\subsection{Results of the MMT/HectoSpec spectroscopy in Cepheus}

First and foremost we find that essentially every target star located in 
the colour-colour plane clearly above the main stellar 
locus is confirmed as an emission line star.  Altogether 29 objects are
confirmed as having H$\alpha$ in emission, with 6 or 7 lying on or just
below the upper bound of the highly populated region (green encircled points
in Fig.~\ref{hec_cep}).  The one object not encircled in green, at $(r' - i')
\simeq 1.1$ and $(r' - H\alpha ) \simeq 1.5$, is probably an emission line
object also, but remains ambiguous because it is very faint and its spectrum
is correspondingly noisy.  There is a still larger group, numbering 47, of 
probable dMe stars.  They are `probable' because the sky subtraction may have 
left a false residue of H$\alpha$ emission.  The range of H$\alpha$ emission 
equivalent widths in this group is from a few up to 10--20\AA .  Quite 
plausibly, most of these objects lie mixed in with non-emission dwarf M stars 
(of which there are 90 or more).  The one `probable dMe' stars (at 
$(r' - i') \simeq 2.50$, $(r' - H\alpha) \simeq 1.3$) above the main locus can 
be viewed as dMe with the greatest confidence because time-variable H$\alpha$ 
emission is characteristic of dMe stars -- presumably at the time of the IPHAS 
imaging, its H$\alpha$ emission was brighter than 9 months later, at the time 
of the spectroscopy.  

The combination of moderate spectral resolution and short exposure 
time has meant that many of the more routine objects, without H$\alpha$ 
in emission or in marked absorption, are more challenging to sort
into spectral classes.  This large group of 214 objects will be dominated by 
late-A to mid K stars, but will also include some non-emission OB stars.  The 
spectra of a further 89 stars are so faint and noisy that no comment can be
made about them.  The stand-out objects towards the lower boundary of 
the main stellar locus in Fig.~\ref{hec_cep} are the stars with H$\alpha$ 
strongly in absorption.  There are 15 of these.  Two of them are well-separated
from the main locus at much lower $(r' - i')$ and also much lower 
$(r' - H\alpha )$ -- they are both white dwarfs.  Similarly placed objects
in other IPHAS fields for which we have MMT/hectospec spectra have turned
out to be white dwarfs too.  The remaining thirteen stars with strong 
H$\alpha$ absorption are early A stars.  Tighter classification at this time 
is not feasible.

\begin{table*}
\caption{Positions, magnitudes and colours for the 8 stars, observed using
MMT/hectospec, whose spectra are plotted in Figs.~\ref{rogues_1} and
\ref{rogues_2}.  The $r'$ magnitudes and $(r'-H\alpha)$, $(r'-i')$ colours 
have been taken from the catalogues for the best pair of IPHAS exposures.
At this time, the magnitudes are likely to be correct to within $\pm$0.1,
while the H$\alpha$ EWs are reliable to within $\pm$5~\AA .  The EW sign 
convention is reversed in that a positive value implies net
emission.}
\label{hec_data}
\begin{tabular}{lclrrlrrrr}
\hline
   & IPHAS name/position & \multicolumn{3}{c}{IPHAS photometry} 
   & object & \multicolumn{3}{c}{2MASS colours/magnitudes} & H$\alpha$ EW \\ 
   & J[RA(2000)$+$Dec(2000)] & $r'$ & $r'-i'$ & $r'-H\alpha $ & type
   & $(J - H)$ & $(H - K)$ & $K$ & (\AA ) \\
\hline  
1 & J221734.39$+$611409.2 & 19.5 & 1.55$\pm$0.06 & 1.72$\pm$0.06 & T~Tau star 
  & 1.22$\pm$0.07 & 0.67$\pm$0.05 & 13.39$\pm$0.03 & 190 \\
2 & J221740.30$+$614702.9 & 17.7 & 1.44$\pm$0.01 & 1.42$\pm$0.01 & B[e]/YSO
  & 1.33$\pm$0.04 & 0.98$\pm$0.04 & 10.52$\pm$0.02 & 200 \\
3 & J221411.60$+$612606.7 & 17.5 & 1.15$\pm$0.01 & 0.99$\pm$0.01 & Be/T~Tau 
  & 1.38$\pm$0.08 & 0.92$\pm$0.05 & 12.12$\pm$0.03 & 80 \\
4 & J221427.30$+$612943.6 & 19.5 & 2.01$\pm$0.08 & 1.08$\pm$0.08 & $\sim$M3Ve 
  & 1.09$\pm$0.04 & 0.56$\pm$0.04 & 13.15$\pm$0.02 & 70 \\
5 & J221534.47$+$615725.4 & 17.2 & 0.06$\pm$0.01 & -0.23$\pm$0.01 & white dwarf
  & -- & -- & -- & $-$34 \\
6 & J221345.86$+$614418.6 & 17.5 & 0.53$\pm$0.01 & 0.00$\pm$0.01 & early A star
  & 0.3$\pm$0.2 & $\sim$0.5 & $\sim$15.3 & $-$16 \\
7 & J221822.09$+$614803.8 & 18.7 & 3.39$\pm$0.03 & 0.78$\pm$0.04 & mid-M giant
  & 1.66$\pm$0.04 & 0.62$\pm$0.04 & 8.68$\pm$0.02 & \\
8 & J221619.89$+$612621.1 & 19.5 & 2.97$\pm$0.06 & 0.71$\pm$0.07 & carbon star
  & 2.61$\pm$0.03 & 1.76$\pm$0.03 & 6.41$\pm$0.02 & \\
\hline
\end{tabular}
\end{table*}


We now present a selection of 8 objects and their spectra for more detailed 
discussion.  These are identified in the colour-colour plane shown as 
Fig.~\ref{hec_cep}.  The data on them, given in Table~\ref{hec_data}, 
includes estimates of their $r'$ magnitudes and $(r' - i')$, $(r' - H\alpha )$ 
colours derived from the highest quality IPHAS exposures currently available.  
Note that the colours are, typically, different from those plotted in 
Figs.~\ref{hec_cep} -- this is due to the colour shifts 
applied in combining IPHAS fields before MMT/HectoSpec target selection.  
Since they assist in assigning broad object class, we also include in the 
table 2MASS $(J-H)$, $(H-K)$ colours and $K$ magnitudes.

\begin{figure}
\begin{picture}(0,380)
\put(0,0)
{\includegraphics{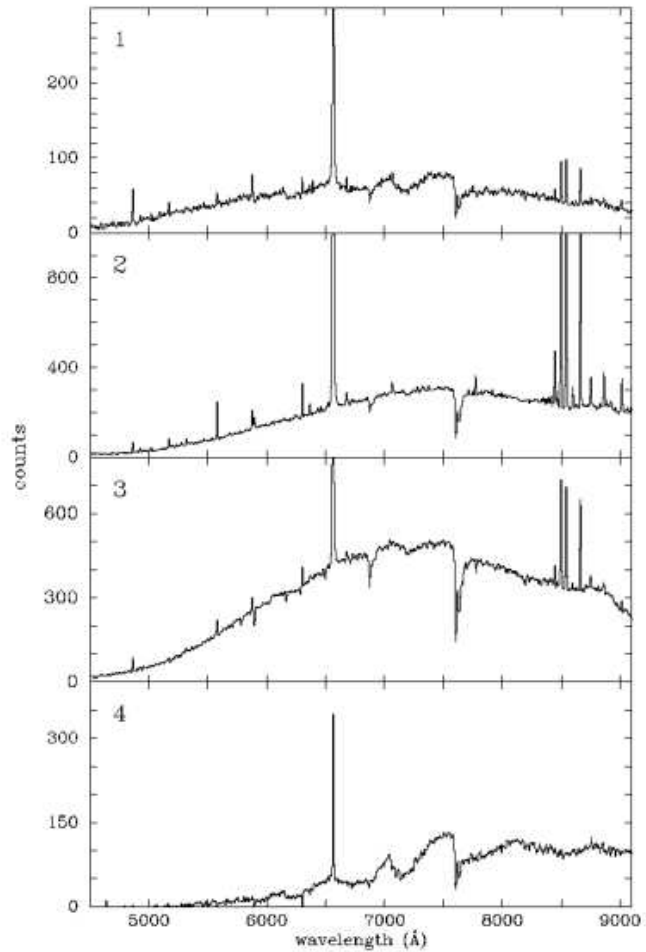}}
\end{picture}
\caption{The MMT/HectoSpec counts spectra of Cepheus targets 1 to 4, as marked 
in Fig.~\ref{hec_cep} and listed in table~\ref{hec_data}.  These data and those
shown in Fig.~\ref{rogues_2} are uncorrected for telluric absorption.
}
\label{rogues_1}
\end{figure}

Stars 1 -- 3 (Fig.~\ref{rogues_1}) are most likely to be young stellar objects 
(YSOs) of Herbig or T~Tau type.  This object class assignment is easiest for 
star 1 since the veiling is not so extreme as to hide the underlying M-star 
spectrum.  Indeed, a comparison between stars 1 and 4 in Fig.~\ref{rogues_1} 
suggests that these objects' M spectral sub-types are 
likely to be very similar. The sky around star 1 has been imaged in all four 
IRAC bands by the {\em Spitzer Space Telescope} First Look Survey. We 
downloaded the calibrated images from the Spitzer Science Archive and carried 
out point source extractions. Star 1 was detected at 3.6, 4.5 and 5.8~$\mu$m, 
with fluxes corresponding to magnitudes of 12.47$\pm$0.11, 12.12$\pm$0.12 and 
11.93$\pm$0.14, respectively. Using the observed IPHAS, 2MASS and IRAC fluxes, 
we find that the SED of Star 1 from the r$'$-band to 5.8~$\mu$m happens to fit 
quite well to a 2000~K blackbody (although the SED must include components
due to a reddened stellar photosphere, an accretion disk and warm dust).

Star 2 has the highest contrast and richest emission line spectrum of the 
three stars -- in this object, even the Na~{\sc i} D lines are thrown into 
emission and some forbidden line emission is present. Star 3 is intermediate 
between 1 and 2, both in terms of the contrast of its emission spectrum, and 
that it is just possible to pick up late-type photospheric absorption against
the continuum (e.g. the blend at 6495~\AA\ strong in G/early-K stars).  The 
NIR colours of all 3 objects imply modest NIR continuum excesses that are not 
out of place for Herbig or T~Tau stars, with star 2 showing the most marked 
excess.  

It seems likely that these stars 1--3 are associated with LDN 1188 
($\ell = 105.7$, $b = +4.2$), a dark cloud less than $2^{\rm o}$ away from the 
well-known Sh~2-140 region of star formation.  This proximity suggests
a similar distance to both regions, which has been given as 910~pc for 
Sh 2-140 by Crampton \& Fisher (1974).  However both these nebulae lie on the 
periphery of Cep~OB2, to which the distance appears to be rather less 
($\sim$600~pc, de Zeeuw et al 1999).  In a study of LDN~1188, Abraham et al 
(1995) reported the discovery of a number of emission line stars in objective 
prism data obtained at Konkoly Observatory at brighter magnitudes than our 
HectoSpec selection.  RNO 140 and RNO 141 (Cohen 1980) are also in this 
neighbourhood.

\begin{figure}
\begin{picture}(0,180)
\put(0,0)
{\includegraphics{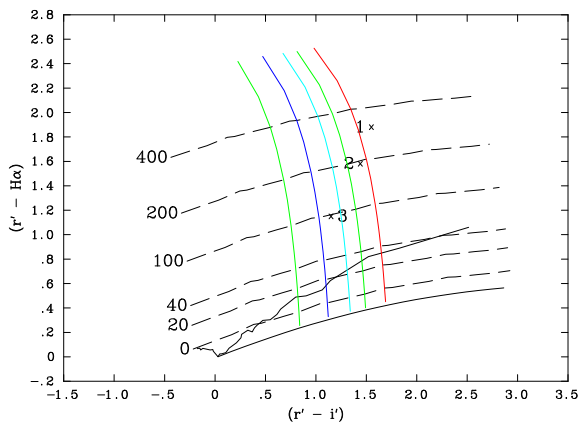}}
\end{picture}
\caption{A diagnostic colour-colour diagram for the Cepheus region stars 
1 to 3 in Table~\ref{hec_data}.  The blue, light blue, (right hand) green and 
red tracks, from Table~\ref{em_colour} and Fig.~\ref{em_tracks}, are reddened 
to $E(B-V) = 2$.  They respectively represent the colour-colour response to 
increasingly strong (narrow) H$\alpha$ line emission  of extreme O stars, 
$\sim$A0 stars, optically-thick accretion disks and $\sim$G2 stars.  For 
contrast, the track
for optically thick disk accretion at $E(B-V) = 1$ (left hand green line)
is also shown.  The horizontal dashed lines are lines of constant H$\alpha$
equivalent width.  The numbers 1--3 mark the positions of stars 1--3: note
that their $(r' - H\alpha)$ colours have been incremented by 0.16 (cf
Fig.~\ref{7012_cc}) to place them in the simulation colour domain.  The 
required correction in $(r' - i')$ will be significantly smaller.
}
\label{rogues_com}
\end{figure}

Star 4 can be presumed to be in the foreground with respect to LDN 1188 and 
stars 1--3.  Hence its $JHK$ colours, combined with its spectral type can be 
used to estimate a minimum interstellar extinction towards LDN 1188.  This 
works out at $E(B-V) \simeq 1.3$ or $A_V \simeq 4$ (for $R = 3.1$, and
using data from Bessell \& Brett 1988).  We can now see if this marries 
up with the implications of the IPHAS colours of stars 1 -- 3 by comparing 
their catalogue values with reddened synthetic estimates.   This is 
accomplished via Fig.~\ref{rogues_com} in which the IPHAS colours for stars 1 
to 3 (Table~\ref{hec_data}) are compared with synthetic tracks 
(Table~\ref{em_colour}).  

If star 2, with its generally high contrast emission line spectrum is a Herbig
or T Tau star with an accretion dominated SED, its reddening would correspond 
to $E(B-V) \simeq 2$: for, in Fig.~\ref{rogues_com}, it lies just to 
the right of the track for $F_\lambda \propto \lambda^{-2.3}$ and $E(B-V)=2$.  
A similar, or somewhat lower, reddening would appear plausible for star 1 in 
that its optical SED should be somewhat redder, intrinsically, than that of 
star 2.  The intrinsic optical SED of star 3 should be intermediate between 
stars 1 and 2 (given the marginal detection of late-type photospheric 
absorption), and yet it is observed to be `bluer' than either.  The highest 
likely reddening of star 3 is $E(B-V) \sim 1.5$: this reddening would apply 
in the limiting case of an accretion-dominated SED, where the contribution of 
the G/K star is small.

On the basis of its $(r'-i')$ colour and spectral type, and after 
correction for its H$\alpha$ emission, star 4 would be assigned 
$E(B-V) \simeq 1$.  This is a somewhat lower estimate than the estimate based 
on 2MASS NIR colours ($E(B-V) \simeq 1.3$) but not so large a discrepancy that 
either the NIR or optical photometrically calibration must be called into 
question.  In conclusion, we find that the order of increasing reddening 
appears to be: star 4 in the foreground, star 3, and then star 1 and star 2, 
spanning the range $1 \lesssim E(B-V) \lesssim 2$.  The Schlegel et al (1998) 
Galactic reddening maps indicate maximal reddenings of $E(B-V) \sim 2.5$ for 
this part of the Plane.  We have rough consistency and a first indication of 
patchy reddening toward the young objects in the vicinity of LDN~1188 .  If 
half the $r'$ flux of star 1 is attributed to an $\sim$M3 stellar photosphere, 
and $A_V \sim 6$, one may deduce a stellar radius of around 4 times the M3V 
main sequence radius, for a distance of $\sim 600$~pc.  

\begin{figure}
\begin{picture}(0,380)
\put(0,0)
{\includegraphics{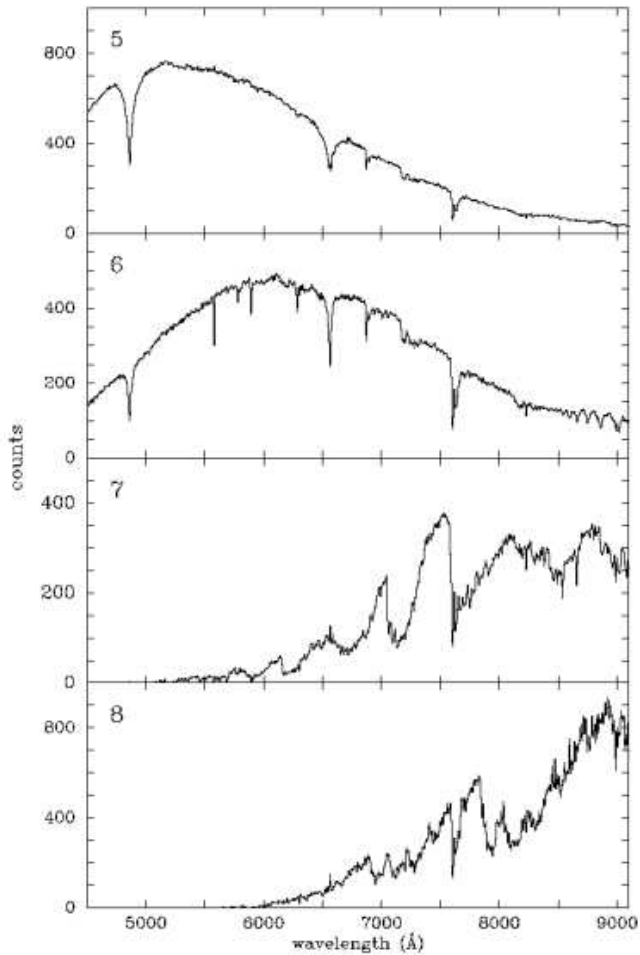}}
\end{picture}
\caption{The MMT/HectoSpec counts spectra of Cepheus targets 5 to 8, as marked 
in Fig.~\ref{hec_cep} and listed in Table~\ref{hec_data}.  
}
\label{rogues_2}
\end{figure}

The IPHAS colours for stars 2 and 3, picked out in Fig.~\ref{rogues_com}, are 
broadly consistent with their observed H$\alpha$ EWs (200 \AA\ and 80~\AA ), 
in that their colours `predict' EWs of $\sim$200 \AA\ and $\sim$100~\AA ,
respectively.  Only star 1 is discrepant in that its observed EW 
($\sim$190~\AA ) is distinctly low compared with the implications of its IPHAS 
colours (suggesting an enormous EW of about 250~\AA ).  Given that H$\alpha$ 
EW is well known to be a time-variable quantity in most classes of emission 
line object, consistency for 2 out of 3 objects is acceptable.  

The spectra of 4 non emission line objects, star 5--8, are shown in 
Fig.~\ref{rogues_2}.  These draw attention to what can be called H$\alpha$ 
{\em deficit} positions in the IPHAS colour-colour plane (cf 
Fig.~\ref{hec_cep}).  At the blue end, objects 5 and 6 are examples of 
extreme and strong H$\alpha$ absorption objects: respectively a white dwarf 
and an early A star.  At the red end, objects 7 and 8 are respectively a 
normal, somewhat reddened mid-M giant, while object 8 has the distinctive CN 
band structure of a carbon star in its spectrum.  Object 7 is typical of the 
stars populating the red end of the giant strip in the IPHAS plane.  The NIR 
colours and rough spectral type ($\sim$M4III) suggest a reddening corresponding
to $E(B-V)$ of about 2.  Carbon stars like object 8 will usually fall below 
the red giant strip (lacking the TiO bands that, in M-type spectra, lead to 
the seeming flux maximum in the H$\alpha$ region).   In terms of its NIR 
colours, star 8 is a more extreme object than the suspected carbon star 
mentioned at the end of section 5 (although reddening gives both objects more 
extreme $(J-H, H-K)$ colours than seen in the local Galactic carbon star 
sample of Claussen et al, 1987).  Indeed Object 8 appears to be a reddened 
($A_V \sim 5$) version of 2MASSI J0326599+143957, described by Liebert et al 
(2000) as a luminous, very cool, late N type carbon star.

In summary, our early MMT/Hectospec observations of this field in Cepheus 
sample a diverse range of objects. Sources lying clearly above the stellar 
locus in the $(r' -H\alpha , r' - i')$ colour-colour plane have indeed been 
confirmed as true emission line objects with H$\alpha$ EWs ranging from a 
few to 200~\AA . Many of them are likely to be Herbig or T~Tau stars.
In addition several H$\alpha$ deficit sources have been identified.
Multi-object spectroscopic follow-up will remain a key part of our efforts 
to mine the IPHAS database.

\section{IPHAS opportunities for H$\alpha$ imaging}

\begin{figure*}
\begin{picture}(0,350)
\put(0,0)
{\includegraphics{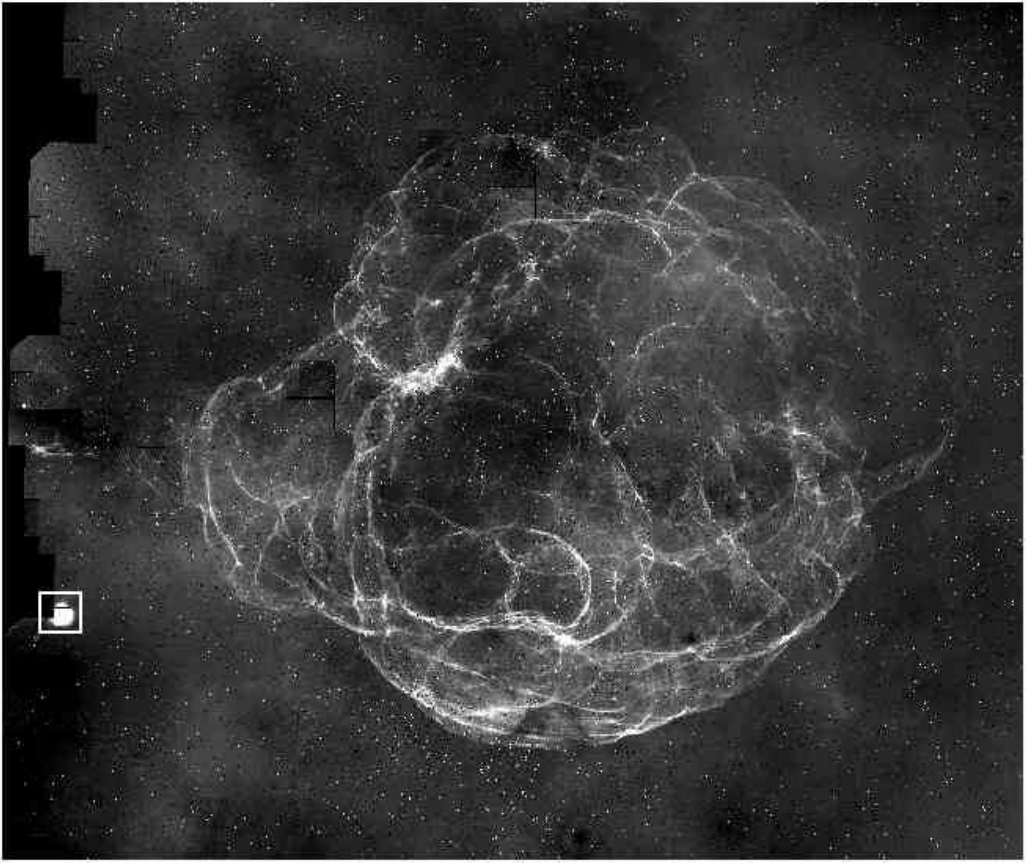}}
\end{picture}
\caption{The continuum-subtracted H$\alpha$ image mosaic of the extremely
large angular size supernova remnant, S~147.  The frame size is roughly 
5 degrees across, and 4 degrees down.  The extent of the SNR is given by 
Van den Bergh et al (1973) as 200 x 180 arcmin$^2$.  N is up and E to the 
left.  The H~{\sc ii} region, Sh~2-242 is picked out near the eastern edge
of the image by the white box (see Fig.~\ref{Sh2-242}).
}
\label{S147_image}
\end{figure*}

The power of the survey for detecting stellar H$\alpha$ emission has already
been described. A second target for the survey are the spatially-resolved 
emission-line nebulae. These nebulae are -- like emission-line stars -- 
associated with early and late phases of stellar evolution, and indicate 
ionization of circumstellar gas, in the form of H~{\sc ii} regions, planetary 
nebulae and supernova remnants.  Already, in the southern hemisphere, the UKST 
SHS has shown the remarkable incompleteness of existing catalogues by doubling 
the number of known planetary nebulae (Parker et al 2003).

The Galactic Plane shows ubiquitous diffuse H$\alpha$ emission as well as
reflection nebulae, which need to be distinguished from circumstellar
ionized nebulae. Since reflection nebulae are the product of continuum 
scattering, they can be removed by comparison of H$\alpha$ and $r'$ images. 
Diffuse H$\alpha$ emission occurs on large size scales (10 arcminutes to 
degrees) and lacks the usual symmetry of circumstellar nebulae. 

Imaging of extended nebulae requires well behaved background on the CCDs making
up the WFC. Because of this, the preferred image properties for nebular
studies are very different from those for point source extraction: high 
background is acceptable for the latter, while poor seeing is acceptable 
for the former.  Small nebulae covered well within a single CCD do not require 
special reduction. Each object will usually have been covered in at least two 
pointings by the time the survey is complete, giving improved S/N. 
For larger objects, mosaics need to be made.  Subtracting an $r'$ frame 
removes stars: for small areas this can be done using PSF matching techniques, 
but this is very computationally intensive -- such that for larger fields, a 
direct subtraction is used which typically leaves larger residuals.

The limitations of the technique are largely due to background variations.
The observations are generally taken in grey and bright time. Different fields
will therefore present with very different background sky levels.  The sky 
subtracts fairly well in an $H\alpha - r'$ image (unless the background is 
variable under non-photometric conditions), but if there is smooth extended 
H$\alpha$ emission over large angular scales, its contribution is currently 
not separately determined from the sky contribution.  Internal reflections are 
seen in some images, from bright stars. In some locations, a bright star just 
outside the field of view gives a flare-like feature on the edge of a nearby 
frame.

A photometric calibration is determined for point sources, but not for
extended emission. To correctly calibrate emission nebulae, assuming
the continuum background is fully subtracted, the filter response curve needs 
to be precisely known and needs to be stable over the likely Doppler 
wavelength shifts (Ruffle et al 2004).  [N~{\sc ii}] will also intrude into 
the flux. Currently, calibration is best performed using known planetary 
nebulae located in the imaged area. Note that stellar H$\alpha$ sources cannot 
be used for calibration as their line flux tends to be time variable.

\subsection{A supernova remnant} 

As an example of the possiblities regarding extended nebulae, we present an
image of the supernova remnant S~147 (Shajn 147, or Simeis 147, in full -- 
not Sh 2-147, with which it is confused in SIMBAD).  This is a 
near-perfect remnant of an approximately spherical shape, showing a typical 
filamentary structure.  It is positioned just overlapping the anti-galactic 
centre, its own centre being at $\ell \simeq 180.1^{\rm o}$, 
$b \simeq -1.6^{\rm o}$.
But due to its large extent, spanning several degrees, only photographic 
images have been published so far (Van den Bergh, Marscher \& Terzian, 1973). 
See also the 24th March 2005 `Astronomy Picture of the Day' due to R. Gendler
(http://antwrp.gsfc.nasa.gov/apod/ap050324.html).

An IPHAS image of S~147 was produced by combining approximately 250 
pointings.   The pipeline mosaicing procedure was found to be inadequate for 
combining many fields taken under widely varying conditions. We therefore used 
the pipeline only to produce reduced images of the individual CCDs for each 
pointing.  Then, for each image, we subtracted the $r'$ image from the 
H$\alpha$ data and smoothed to a pixel size of 5 arcsec (a binning factor of 
15). The background per image was approximately nulled by subtraction of the 
median: in fields with bright and extended emission this required manual 
selection of areas for background definition -- otherwise the median over all 
four CCDs of one pointing was used.  One corner of CCD no. 3 is affected by 
scattered light in conditions of bright Moon light (amplified if cirrus was 
present): this corner was always blanked out. All CCDs tend to show a gradient 
along the long axis of about 1 ADU in the subtracted image: this could be due 
to a charge-transfer efficiency limitation, but its exact cause is not known. 
It does not subtract automatically because the $r'$ image is scaled first -- 
accordingly, as a final step, this remaining gradient was subtracted from all 
images.

The resulting images were combined in a single mosaic using the
Virtual Observatory software package Montage. This involved regridding each
frame, and determining background corrections by comparing areas in common
between different images.  We finally produced an image covering 25 square
degrees. We note that extended emission on scales of a degree or more may
not be well represented, as it can be affected by the background subtraction
procedure. However, the filamentary structures are very well recovered.  The
background fitting was found to be insufficiently constrained at the outer
edges of the imaged area, leaving some negative areas.  To deal with this, a 
linear gradient was fitted to the background in empty regions of the full 
field, and subtracted.

\begin{figure}
\begin{picture}(0,200)
\put(0,0)
{\includegraphics{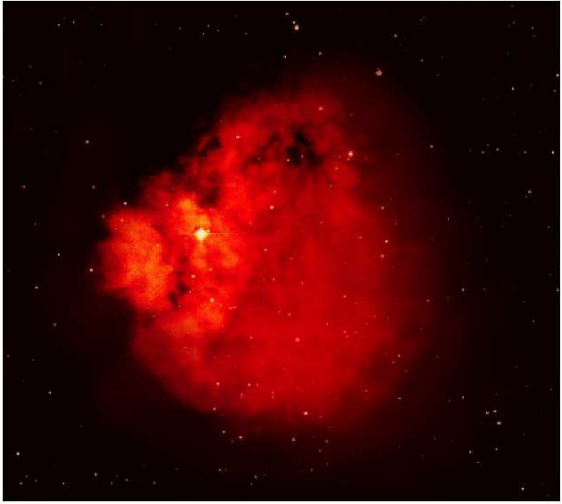}}
\end{picture}
\caption{The H{\sc ii} region, Sh 2-242, picked out on the eastern edge of 
Fig.~\ref{S147_image}, shown at full resolution.  This is a false colour 
composite of the images in the three filters. The imaged area is roughly 
10 x 10 arcmin$^2$.  N is up and E to the left.
}
\label{Sh2-242}
\end{figure}

The image of S~147 is shown as Fig.~\ref{S147_image}.  A bigger version
is to be found in the ING Newletter article describing IPHAS (Drew et al 2005).
The improvement is very marked in comparison with the photographic image 
shown by Van den Bergh et al (1973): the huge increase in dynamic range brings 
with it a subtlety of detail missing from the old imagery.  Furthermore, these 
data allow smaller areas to be imaged to much higher resolution.   But even at 
this 5~arcsec rebinning the structure is clear. We note that a blow-out is 
obvious on two sides of the remnant: left and right (to the E and W).  The 
full extent of S~147 is essentially as reported by Van den Bergh et al, at a 
mean diameter of just over 3$^{\rm o}$ -- several times that of the Moon.

At the eastern edge of the image mosaic, and south of centre, lies another
prominent but much more compact nebulosity, Sh~2-242.  This very bright smudge 
in Fig.~\ref{S147_image} has a mean H$\alpha$ diameter of 8.0 arcmin.  As a 
contrast to the very large scale structure of S~147, we show as 
Fig.~\ref{Sh2-242}, a full resolution image of this H~{\sc ii} region.  

\section{Summarising discussion} 

The main aim of this paper has been to introduce IPHAS, the INT Photometric
H$\alpha$ Survey of the Northern Galactic Plane.  By the beginning of 2005, 
55\% of the imaging observations had been obtained.  It is expected that 
the survey will reach completion in 2006.

Given that previous surveying of the northern Galactic Plane for emission 
line objects rarely reached deeper than $V \sim 13$ (see KW99), while the 
sensitivity limit of IPHAS is $r' \sim 20$, there is no doubt that a huge 
domain is being opened up for exploration for the first time.   It is 
difficult to predict the numbers of emission line objects that will be
discovered, even now with the survey in progress, because the distribution of 
such objects along the Galactic Plane is extremely uneven (see KW99).  The 
small number of IPHAS fields discussed in this paper fit in with this 
impression: no emission line objects were evident in the colour-colour plane 
for the Taurus field (2540/2540o in section 5 -- but a very modest example was 
found on closer examination); a handful were evident in the Aquila fields 
(section 4), and upwards of 20, all fainter than $r' = 17$, were picked up in 
the MMT/HectoSpec pointing toward LDN~1188 in Cepheus (section 6).  Crudely 
averaging the experience to date, it is likely IPHAS will uncover around 10 
emission line objects per square degree in the range $13 < r' < 20$ (roughly 
2 to 3 per IPHAS field), and hence no less than $\sim$20,000 altogether.  

Through its use of both narrow-band H$\alpha$ and broadband $r'$ and $i'$ 
CCD photometry, IPHAS has the capability to pick out H$\alpha$ deficit 
objects -- a possibility typically beyond objective prism spectroscopy, the 
traditional tool of emission line star hunting.  For example, unreddened white 
dwarfs are easily spotted as blue objects with small or negative 
$(r' -H\alpha)$, separated from the main stellar locus in the IPHAS
$(r' - H\alpha, r' - i')$ plane.  It is a reasonable guess that over a 
thousand will be discovered in IPHAS data, just as have been discovered in a 
comparable sky area around the north Galactic cap via the Sloan survey 
(Kleinman et al 2004).  Very red H$\alpha$ deficit objects can be either 
brown dwarfs or carbon stars -- classes of star that both lack the TiO band 
absorption responsible for raising $(r' - H\alpha)$ in normal M stars -- or 
they can be very reddened, rare examples of late-type supergiants.  Two 
carbon stars (one probable, the other confirmed) have been reported here.

A further role for IPHAS is that it can help trace the way in which the 
stellar populations making up the Galactic Plane vary across the northern sky.
It has been shown here that not all sightlines look the same in IPHAS 
colours: of particular note is the sharp contrast between the red-giant 
deficient Taurus field ($\ell = 181.7^{\rm o}$, section 5), in the Galactic 
Anticentre region, versus the Aquila fields ($\ell \simeq 33^{\rm o}$) with 
their prominent giant-star populations, sampling the inner Galaxy.   Before 
now, rather little has been known about the far reaches of the Milky Way 
outside the Solar Circle.  For example, the recent study of 
Galactic spiral structure by Russeil (2003) reaches only to $\sim 6$~kpc, 
outward from the Sun.  The sensitivity of IPHAS is more than adequate for 
extending our knowledge much further: even relatively humble A0--A3~V stars 
-- easily picked out around what has been dubbed the early-A reddening line 
in the IPHAS colour-colour plane -- are potentially accessible out to 
distances of $\sim$20~kpc in the direction of the lightly reddened 
Galactic Anticentre.

We have laid out the character of the $(r' - H\alpha, r' - i')$ 
colour-colour plane that is unique to this survey, and have established a 
grid of simulated colours that will be of use in the analysis of IPHAS 
observations.  In doing this, we have inevitably identified items of work 
for the future.  For example, a job remains to be done to achieve as good a
quantitative match between simulations and observations of M stars as is 
achieved for earlier spectral types.  A related problem has been noted by 
those developing Sloan Digital Sky Survey photometric calibrations (Smith et 
al 2002).  The reward for solving this problem will be the chance to fully
realise the potential of IPHAS as an unparalleled resource for the 
statistical analysis of M-dwarf activity.   Another task is to gather a
a range of early A star spectra that can be used to achieve better definition 
of the early-A reddening line and its dependences.  As specified here, it
is unlikely to be more than $\sim$0.02 magnitudes from its correct
registration in $(r' - H\alpha )$.

Bigger issues to be dealt with are that the photometric calibration of IPHAS 
data will need to be made uniform across the survey, and that a proper 
definition of H$\alpha$ magnitude zero point (referred to Vega) is needed that 
can be applied within the pipeline processing.  Until these calibration 
requirements have been met, it will remain necessary to derive colour offsets 
between observed and simulated data independently for every IPHAS field.  
Typically, offsets in $(r' - i')$ will be small -- but for $(r' - H\alpha)$, 
the differences between catalogue and simulated colours may vary from 
$\sim$0.1 up to $\sim$0.2 magnitudes.  In most cases it is easy to gauge the 
required offset by matching the synthesised unreddened main sequence track and 
early-A reddening line to the upper and lower boundaries of the main stellar 
locus in the colour-colour plane.

Simulations have also been performed that show how the equivalent width 
threshold for the detection of H$\alpha$ emission will change with observed
$(r' - i')$ colour (section 3.2 and Fig.~\ref{em_tracks}).  The photometric
accuracy of IPHAS is such that $(r' - H\alpha)$ differences of 0.05--0.1 are 
significant down to $r' \sim 20$.  Re-expressed in terms of an H$\alpha$ 
equivalent width, and for magnitudes brighter than $\sim$19, this corresponds 
to a threshold emission EW of roughly 5~\AA . 

It has been shown that the typical morphology of the main stellar locus in the 
colour-colour plane permits the selection of candidate emission line stars 
presenting with threshold H$\alpha$ emission at low $(r' - i')$ only.  
In practise the IPHAS bright magnitude limit ($r' \sim 13$) has the effect of 
all but eliminating normal stars with colours bluer than $(r' - i') \sim 0.5$ 
from the point source catalogue (note Figs.~\ref{aquila_cc}, \ref{2540_cc} 
and \ref{cepheus_cc}) -- with the consequence that lightly-reddened 
subluminous accreting objects, with or without H$\alpha$ emission, will 
usually lie comfortably outside the main stellar locus.  IPHAS can therefore 
be used straightforwardly to identify all such objects (in addition to many 
non-interacting white dwarfs, found as `deficit' objects).  

The threshold for H$\alpha$ emission high-confidence detection rises from 
$\sim 10$~\AA\ equivalent width at $(r' - i') \sim 1$ up to $\sim 50$~\AA\
at $(r' - i') \sim 2.5$ -- beyond this the colour-colour plane typically
becomes sparsely populated again.  This has important implications for the 
detectability of T~Tau and other young emission line objects: based on 
equivalent width data collected by Reipurth, Pedrosa \& Lago (1996, their
Table 1 and Fig. 10), one-third to a half of such objects would be immediately 
identifiable as emission line objects in the IPHAS database at 
$E(B-V) \simeq 3$.  At lower reddenings the fraction would be higher.
This suggests that the roles for IPHAS with respect to young stellar 
populations are (i) finding extreme examples anywhere (here we have presented 
two with H$\alpha$ emission EWs of $\sim 200$~\AA ), and (ii) picking out new 
associations through the detection of its most active members.  Once a new 
association is identified, a more detailed exploration of catalogued IPHAS 
sources, exploiting the tricks of e.g. apparent magnitude binning, or small 
area searches, would be likely to uncover further candidate emission line 
sources.  

Evolved high mass stars, with H$\alpha$ emission, that find their way into the 
IPHAS database will do so because they are very distant and significantly 
reddened (they are otherwise too bright).  The more extreme, least 
well-understood, and therefore more interesting groups (Wolf-Rayet stars, 
luminous blue variables, B[e] stars, yellow hypergiants) usually present with 
very high EW H$\alpha$ emission ($\sim100$~\AA\ and more) and so will not be 
missed.  The most extreme point-source emission line objects of all,
compact nebulae, are most likely to appear to be relatively blue in 
$(r' - i')$ or even evade detection in the $i'$ band, and will have the
most extreme values of $(r' - H\alpha)$ possible, i.e. not more than
3 to 3.1, in practice. 

This connects naturally to a topic only touched on in this paper --  the 
exploitation of IPHAS in the study of spatially-resolved nebulae (deferred to 
a later paper).  A search for PNe is in progress, in which several tens of 
candidates have been identified and a few have been studied spectroscopically. 
A paper on the study of an intriguing quadrupolar nebula located well-outside 
the Solar Circle is in preparation (Mampaso et al).  Here we have simply drawn 
attention to the flexibility IPHAS presents for the investigation of a wide 
range of spatial scales, ranging from arcseconds to several degrees.

Further power to diagnose either particular object types or complete stellar
populations will come from pooling IPHAS $r'$, $H\alpha$ and $i'$ data with 
data from surveys in different wavebands.  In the future of an astronomy 
conducted via virtual observatories, a survey as comprehensive as IPHAS -- 
with its particular exploitation of narrowband H$\alpha$ data -- will be
a major resource.  Currently there is an obvious synergy with the all-sky NIR 
survey 2MASS, although this is limited to the reddened parts of the northern 
Galactic Plane since 2MASS reaches only to $K \sim 15$.  The gap this leaves 
should soon be plugged by the UKIDSS Galactic Plane Survey, reaching to $K=19$ 
(see http://www.ukidss.org/).  Beyond IPHAS, comprehensive optical surveying 
of the northern Galaxy using linear detectors is still lacking.  However it is 
interesting to note the recent release by SDSS of u'g'r'i'z' data on a number 
of low Galactic latitude fields (Finkbeiner et al 2004).


We finish with a comment on the plans for making IPHAS data available to the 
community.  At the present time there is open access to the reduced images 
held at CASU, from a year after the images have been processed. Accordingly 
the images obtained in 2003 can already be accessed by anyone able to reach 
the CASU web interface (http://apm2.ast.cam.ac.uk/cgi-bin/wfs/dqc.cgi).  
Access immediately after processing is available to those working in the ING 
partner countries: the United Kingdom; Spain and The Netherlands.  The point 
source catalogues are to be released in two stages: in the first half of 2006 
we aim to release as many as are available, calibrated only at the individual 
exposure level (as they have been described here); we intend to follow this up 
with a second release, after the survey is complete, and when a uniform 
calibration has been established across all fields.  We anticipate that the 
final catalogue will contain photometry on around 80 million point sources,
capturing data on roughly 1 for every 1000 stars estimated as existing in the 
northern Milky Way.

\section*{Acknowledgments}
This paper makes use of data from both the Isaac Newton and William Herschel 
Telescopes, operated on the island of La Palma by the Isaac Newton Group in 
the Spanish Observatorio del Roque de los Muchachos of the Instituto de 
Astrofisica de Canarias.   The multi-object spectroscopic observations 
reported here were obtained at the MMT Observatory, a joint facility of the 
Smithsonian Institution and the University of Arizona.  We would like to thank 
the HectoSpec team for their assistance: in particular, Nelson Caldwell and 
Perry Berlind for their help with the data acquisition, and Susan Tokarz for 
the pipeline reduction products.  We also acknowledge 
use of data products from the Two Micron All Sky Survey (2MASS), which is a 
joint project of the University of Massachusetts and the Infrared Processing 
and Analysis Center/California Institute of Technology (funded by the USA's 
National Aeronautics and Space Administration and National Science 
Foundation).  DS acknowledges a Smithsonian Astrophysical Observatory Clay 
Fellowship.

\end{document}